\documentclass[12pt]{article}

\usepackage{amssymb,amsmath, amsfonts, amsthm}
\usepackage{graphicx}
\usepackage{caption}
\usepackage{subcaption}
\usepackage{enumerate}
\usepackage{enumitem}
\usepackage{bbm}
\usepackage{url} 
\usepackage[most]{tcolorbox}
\usepackage{ragged2e}
\usepackage[capposition=top]{floatrow}
\usepackage{booktabs}
\usepackage{adjustbox}
\usepackage{threeparttable}
\usepackage{multirow}
\usepackage{tabularx}
\usepackage{array}
\usepackage{siunitx}
\usepackage{dsfont}
\usepackage[unicode=true,pdfusetitle,
 bookmarks=true,bookmarksnumbered=false,bookmarksopen=false,
 breaklinks=false,pdfborder={0 0 1},colorlinks=true]{hyperref}
\hypersetup{linkcolor=red, citecolor=blue}
\usepackage{natbib}
\newcommand{\blind}{1}
\usepackage{bm}
\usepackage{xcolor}

\newcommand{\E}{\operatorname{E}}

\renewcommand{\P}{\operatorname{P}}

\newtheorem{assumption}{Assumption}[section]
\newtheorem{proposition}{Proposition}[section]

\newcommand{\mc}[1]{\multicolumn{1}{c}{#1}}

\begin{document}

\def\spacingset#1{\renewcommand{\baselinestretch}%
{#1}\small\normalsize} \spacingset{1}

\if1\blind
{
\title{\bf A Machine-Learning-Compatible Omnibus Test for Treatment Effect Heterogeneity\thanks{Elia Lapenta acknowledges funding from the French National Research Agency (ANR) under grant ANR-23-CE26-0008-01. Anthony Strittmatter acknowledges funding from the Swiss National Science Foundation (SNSF) under grant 10006811. The authors are grateful for helpful comments and discussions from Laurent Davezies and Xavier D'Haultf\oe uille and seminar participants at CREST. All remaining errors are ours.}}
  \author{
    Elia Lapenta\\
    University of Exeter and CREST\and
    Anthony Strittmatter\\
    UniDistance Suisse and CREST \and 
    Pedro Vergara Merino\\
    CREST and ENSAE Paris
    }
     \maketitle
} \fi

\if0\blind
{
  \bigskip
  \bigskip
  \bigskip
  \begin{center}
    {\LARGE\bf One-step  nonparametric instrumental regression using
  smoothing splines}
\end{center}
  \medskip
} \fi

\thispagestyle{empty}

\begin{abstract}
This study proposes a formal, computationally efficient nonparametric omnibus test for treatment-effect heterogeneity that is compatible with a broad class of estimators, including modern machine-learning methods. The test is designed for settings in which identification can rely on high-dimensional controls while heterogeneity is assessed with respect to a low-dimensional subset of covariates. We derive the test statistic’s asymptotic null distribution and develop a bootstrap procedure that is efficient because it avoids re-estimating nuisance parameters in each iteration. The testing approach applies to multiple empirical designs, including randomized experiments, selection-on-observables, difference-in-differences, and instrumental-variables settings. Monte Carlo simulations show that the test attains near-nominal size under the null and exhibits good power against heterogeneous alternatives. We further illustrate the procedure using two empirical applications on retirement savings and trade liberalization.
\end{abstract}
\vspace{0.2cm}

\noindent%
\vspace{0.2cm} 

\noindent%

\clearpage
\spacingset{1.1} 
\section{Introduction}
Understanding whether treatment effects vary is central to policy evaluation because average effects may conceal differences in how policies, programs, or interventions affect different groups. Heterogeneous effects are relevant in many fields, including economics, political science, and health sciences, where researchers routinely investigate whether an intervention benefits some groups more than others. For example, one may want to assess whether a job-training program improves employment differently across skill levels, whether the effectiveness of a medical treatment varies by the age of patients, or whether an educational intervention influences learning gains differently across socioeconomic groups. In all these cases, a valid hypothesis test for treatment effect heterogeneity is essential.

In recent years, machine learning (ML) methods have become increasingly popular for uncovering such heterogeneity \citetext{see, e.g., \citealp{athey2019machine} and \citealp{chern2024}}. Compared with traditional econometric approaches such as interacted models or subgroup analysis, ML-based techniques typically require fewer parametric assumptions, offer more flexible modeling with safeguards against overfitting, and can accommodate large sets of control variables, often high-dimensional ones. Many of these methods were originally developed for prediction problems. In many empirical applications, however, researchers are interested in causal parameters that are directly relevant for policy evaluation, such as average treatment effects (ATEs) and their variation across groups. Traditional econometric methods therefore place particular emphasis on identification and valid statistical inference for such quantities. Recent advances in Causal-ML combine the flexibility of ML with the inferential objectives of econometrics by using ML methods to estimate high-dimensional nuisance functions while preserving valid inference for causal effects. Despite this progress, there remains a need for formal procedures that can assess treatment-effect heterogeneity while retaining the flexibility of modern ML methods and that can be readily implemented in practice.

This study contributes to the small but growing literature on detecting systematic treatment effect heterogeneity with ML (see literature review below). It develops a nonparametric omnibus test for treatment-effect heterogeneity that can be combined in a modular way with a wide range of estimators, including modern ML methods. The test provides a single global assessment of whether treatment effects exhibit systematic variation with respect to a chosen low-dimensional set of covariates, while allowing identification to rely on high-dimensional controls. We construct a locally robust test statistic that remains valid under flexible nuisance estimation and pair it with a computationally efficient bootstrap. The bootstrap requires nuisance functions to be estimated only once, avoiding repeated first-stage estimation in each iteration. Together, these features yield a practical tool for large empirical applications and for settings where heterogeneity is of interest but conventional low-dimensional methods are inadequate.

The broad applicability of the procedure is illustrated across several common empirical designs. We first consider heterogeneity in ATEs under conditional unconfoundedness, the standard identifying assumption in observational studies. The same framework naturally covers randomized experiments as a special case in which treatment is independent of covariates and nuisance functions are either known or estimated only to improve precision (for example, \citealp{list2025}, shows that ML-based covariate adjustment in experiments can enhance efficiency without affecting identification). We then extend the approach to difference-in-differences (DiD) designs to test heterogeneity in average treatment effects on the treated (ATT) under conditional parallel trends, accommodating both panel data and repeated cross-sections. Finally, we adapt the method to instrumental-variables settings to test for heterogeneity in local average treatment effects (LATE) when instruments are valid conditional on covariates.

For each design, we derive the corresponding test statistic and establish its asymptotic null distribution. The construction follows Integrated Conditional Moment principles in the spirit of \citet{bierens2016econometric}. We reformulate the conditional null of no treatment-effect heterogeneity as a continuum of unconditional moment restrictions and integrate these moments to obtain a feasible test. Because the limiting distribution is not directly useful for implementation, we introduce a bootstrap approximation that is asymptotically valid under the null. The procedure requires estimating nuisance components only once, which we do using ML methods with cross-fitting. Neyman orthogonality of the statistic ensures local robustness to regularization bias, so small first-stage errors do not affect the asymptotic distribution. Estimating nuisances a single time and relying on simple matrix operations within the bootstrap makes the method computationally efficient.

Simulation results confirm the practical performance of the test. Across a range of sample sizes, covariate dimensions, and ML estimators for the nuisance components, the procedure attains close-to-correct size under the null and exhibits high power against relevant heterogeneous alternatives.

We illustrate the empirical relevance of the procedure using two applications. The first revisits the effect of eligibility for a 401(k) retirement savings plan on household financial assets under conditional unconfoundedness, following \citet{poterba_401k_1995} and \citet{chernozhukov2018double}. The second considers treatment effect heterogeneity in a difference-in-differences setting studying the impact of tariff reductions on corruption along the South Africa--Mozambique trade corridor, building on \citet{sequeira_corruption_2016} and \citet{chang2020double}. Across both applications, the proposed test detects economically meaningful heterogeneity while remaining computationally feasible in settings with flexible ML-based nuisance estimation. 

\textbf{Related Literature.} This study contributes to a growing body of work on testing and estimating heterogeneous treatment effects using semi- and nonparametric tools, often combined with modern ML. We situate our approach within three related strands of the literature: a smaller literature on global inference for systematic heterogeneity, a smaller literature on tests of identical unit-level effects, and a large literature that uses ML under unconfoundedness to estimate heterogeneous treatment effects through conditional average treatment effects (CATEs) in high-dimensional settings.

A first strand develops global procedures for testing whether treatment effects vary systematically with observed covariates. Early work such as \citet{crump2008nonparametric} proposes nonparametric homogeneity tests across covariate-defined subpopulations, but these methods are mainly designed for low-dimensional covariates and classical estimators. More recent contributions combine global heterogeneity assessment with ML methods. \citet{chernozhukov2023fisherschultzlecturegenericmachine} develop a general sample-splitting framework for inference on low-dimensional summaries of treatment-effect variation, including best-linear predictors and sorted effects. In this framework, heterogeneity is assessed through an ML-generated score, either by testing whether the score is linearly related to treatment-effect variation or by comparing average effects across groups ranked by the score. Building on related ideas, \citet{Imai2025a} propose a nonparametric heterogeneity test for randomized experiments based on sorted effects. Our approach differs from these contributions in three main respects. First, our objective is to test directly whether treatment effects vary systematically with a chosen set of covariates, without requiring heterogeneity to be summarized by a linear projection or a small number of ranked groups. Second, while these approaches focus primarily on randomized experiments or settings under unconfoundedness, our framework extends to several empirical designs, including difference-in-differences and instrumental-variables settings. Third, our procedure remains computationally efficient because the bootstrap avoids repeated re-estimation of nuisance components.

A second strand of research studies randomization-based tests built for the sharp hypothesis of identical unit-level effects. Key contributions include \citet{heck97} and \citet{ding2016randomization,ding2019decomposing}. These tests are designed to detect any deviation from constant individual effects, including purely idiosyncratic variation. In contrast, our target is a weaker but empirically relevant null, namely the absence of explainable heterogeneity with respect to observed covariates. This distinction matters in applications where the goal is to determine whether effects differ across observable groups in a learnable and actionable way, rather than whether unit-level heterogeneity is driven by potentially idiosyncratic and unpredictable variation.

A third strand examines effect heterogeneity under unconfoundedness using flexible ML methods, focusing primarily on the estimation and inference of CATEs and related functionals in high-dimensional settings. This literature includes tree-based methods such as causal trees and causal forests \citep{ath2016,wager2018estimation,ath2019,Fried21,lech2022}, debiased, orthogonalized, and partialling-out approaches for high-dimensional CATE inference \citep{semenova2021debiased,chernozhukov2018,chernozhukov2021,fan2022,lech2019,semenova2017,nekipelov2022,Foster2019,kennedy2023towards,Nie2020,scheidegger2026inference}, approaches imposing additional structure on effect modifiers, such as sparsity in targeted undersmoothing \citep{Hansen2017}, and Bayesian models for CATE heterogeneity based on hierarchical or prior-driven structure \citep{green2012,tad2016,hahn2020}. Taken together, this strand provides powerful tools for estimating and interpreting heterogeneous effects, primarily under unconfoundedness, but it does not provide a general omnibus test of whether systematic heterogeneity is present with respect to observed covariates.

Finally, our paper is related to \citet{escanciano2026kernel}, who develops a specification test for moment restrictions using Reproducing Kernel Hilbert Spaces, in settings where the null hypothesis is characterized in terms of high-dimensional nuisance parameters.\footnote{This paper was presented under the title "Kernel-Based Specification Testing with High-Dimensional Nuisance Parameters" at the ISNPS Conference in Thessaloniki, June 2026. A written version of the paper was not yet available at the time of writing.} Our contribution differs from his along three dimensions. First, we focus on a specific class of moment restrictions arising from treatment-effect heterogeneity, rather than on conditional moment restrictions more broadly. Second, we establish an expansion of the empirical process underlying our test statistic that is uniform with respect to the indexing parameter $t$, which underlies our asymptotic results. Third, whereas \citet{escanciano2026kernel} obtains critical values through a multiplier bootstrap based on the influence-function representation of the test statistic, we propose a weighted bootstrap that departs from the multiplier scheme; in our empirical applications, this construction avoids the computation of certain terms that would otherwise complicate implementation.

\textbf{Outline.} Section \ref{sec:intuition} provides an intuitive overview of the proposed testing framework and its motivation. Section \ref{sec: framework} develops the test for heterogeneity in ATEs under unconfoundedness, including the locally robust statistic and bootstrap procedure. Section \ref{sec: beyond} extends the framework beyond unconfoundedness to difference-in-differences designs with panel and repeated cross-section data, as well as to LATEs identified through instrumental variables. Section \ref{sec: simulations did} reports simulation evidence on size and power. Section \ref{sec: empirical applications} presents two empirical applications illustrating the proposed procedure in observational and difference-in-differences settings. The final section concludes.

\section{Intuition and Overview of the Test}\label{sec:intuition}

Researchers are often interested not only in whether an intervention affects outcomes on average, but also in whether its effects differ across groups. The potential-outcomes framework provides a natural way to formalize this question. Let $D_i\in\{0,1\}$ denote a binary treatment indicator, where $D_i=1$ if unit $i$ receives the treatment and $D_i=0$ otherwise. Depending on the application, the treatment may represent participation in a job-training program, eligibility for a tax-advantaged savings plan, access to a microcredit program, or exposure to a policy reform. Let $Y_i(d)$ denote the potential outcome under treatment status $d\in\{0,1\}$, such as employment, household income, savings, or firm performance. Thus, $Y_i(1)$ is the outcome that would be observed if unit $i$ received the treatment, whereas $Y_i(0)$ is the outcome that would be observed if the same unit did not receive the treatment. The difference $Y_i(1)-Y_i(0)$ is the individual treatment effect. Let $X_i$ denote a vector of observed pre-treatment covariates, such as age, education, earnings and employment histories, household characteristics, or pre-treatment measures of business performance. For notational convenience, let $(Y(1),Y(0),D,X)$ denote generic population random variables corresponding to the unit-level quantities $(Y_i(1),Y_i(0),D_i,X_i)$.

A fundamental challenge of causal inference is that only one of the two potential outcomes is observed for any given unit. Individual treatment effects are therefore not directly observed. Nevertheless, averages of these individual effects can be identified from observed data under appropriate assumptions. A central example is the ATE,
$\delta_0=\operatorname{E}\{Y(1)-Y(0)\}$,
which measures the average impact of the intervention in the population. For example, in an evaluation of a job-training program, $\delta_0$ captures the average change in employment that would occur if everyone in the population participated in the program rather than if no one did.

In many empirical applications, researchers face two related but distinct tasks. The first is identification: estimating causal effects may require adjusting for a rich set of covariates. The second is interpretation of treatment-effect heterogeneity: substantive interest is often focused on whether effects differ across a smaller set of policy-relevant characteristics. We therefore distinguish between the full covariate vector $X$, which is used for identification, and the lower-dimensional subvector $X_c\subseteq X$, which contains the covariates with respect to which heterogeneity is assessed.

This distinction arises naturally in practice. In evaluations of job-training programs, identification may require controlling for detailed labor-market histories, demographic characteristics, regional indicators, and prior earnings, while the main question may be whether effects differ by age or educational attainment. In studies of medical treatments, adjustment may require many clinical and behavioral variables, while interest may center on whether treatment effects vary with baseline health status. In educational applications, researchers may control for a rich set of student, family, and school characteristics, while examining whether an intervention has different effects by gender, migration background, or socioeconomic status. In each case, the full vector $X$ may be high-dimensional, whereas the heterogeneity question concerns a lower-dimensional set of covariates $X_c$.

Treatment effects may also be summarized conditionally on observed characteristics. We write
\[
\delta(X):=\operatorname{E}\{Y(1)-Y(0)\mid X\}
\]
for the CATE conditional on the full covariate vector $X$. Since our test focuses on heterogeneity with respect to the lower-dimensional covariate vector $X_c$, we also define
\[
\delta_c(X_c):=\operatorname{E}\{Y(1)-Y(0)\mid X_c\}.
\]
Thus, $\delta(X)$ describes treatment-effect variation with respect to the full set of covariates, whereas $\delta_c(X_c)$ describes treatment-effect variation with respect to the covariates used in the heterogeneity test.

Our objective is to test whether treatment effects vary systematically with $X_c$. Formally, for the ATE, we consider the null hypothesis
\begin{gather}\label{eq: null hypothesis}
    \mathcal{H}_0:
    \delta_c(X_c)
    =
    \delta_0
    \text{ for some fixed constant }\delta_0\in\mathbb{R}\,,\\
    \text{ versus }\mathcal{H}_1:\mathcal{H}_0^c\,, \nonumber
\end{gather}
where $\mathcal{H}_0^c$ denotes the logical complement of $\mathcal{H}_0$. Under the null hypothesis, the CATE indexed by $X_c$ is constant and equals the ATE, $\delta_0$. The null does not require $\delta_0$ to be zero, nor does it require all individual treatment effects to be identical. Rather, it states that the conditional average effect with respect to $X_c$ is constant. Under the alternative, the conditional average treatment effect varies systematically with at least one component of $X_c$.

 This null is the analogue, for the chosen heterogeneity covariates $X_c$, of the treatment-effect homogeneity null considered by \citet{crump2008nonparametric}, under which CATEs are identical across covariate-defined subpopulations. In contrast to their setting, our procedure allows identification to rely on a potentially high-dimensional vector of controls $X$, with the required nuisance functions estimated flexibly using ML methods.

A common alternative way to study treatment-effect heterogeneity is to estimate the CATEs, because they describe how treatment effects differ by observed characteristics and can therefore help identify which groups benefit more or less from an intervention. Modern examples include tree- and forest-based methods, as well as debiased and orthogonalized approaches for high-dimensional settings \citep[e.g.,][]{ath2016,wager2018estimation,ath2019,semenova2021debiased,fan2022}. Our goal is different. Estimating the full conditional effect function is more ambitious than testing whether systematic heterogeneity is present. In addition, inference on CATEs is often used to test whether the treatment effect is zero at particular covariate values, for example
\[
    \mathcal{H}_0^{\mathrm{zero}}(x):
    \operatorname{E}\{Y(1)-Y(0)\mid X=x\}=0.
\]
This is not a general test of heterogeneity. A treatment effect may be constant and nonzero for all groups. In that case, a zero-effect null can be rejected even though there is no treatment-effect heterogeneity. Our test instead asks whether the conditional average effect differs across values of $X_c$, while allowing the average effect itself to be nonzero.

Recent ML-based approaches also provide formal tests for treatment-effect heterogeneity through summaries of treatment-effect variation \citep{chernozhukov2023fisherschultzlecturegenericmachine,Imai2025b}. Best-linear-predictor (BLP) methods first use ML to construct a score, denoted by $S(X)$, that is intended to predict the CATE $\delta(X)$. This score is not the true CATE, which is unobserved, but a first-stage proxy for heterogeneity. The BLP approach then asks whether actual treatment effects are systematically related to this score by considering the population projection
\[
    \operatorname{BLP}\{\delta(X)\mid S(X)\}
    =
    \beta_1+\beta_2\{S(X)-\operatorname{E}[S(X)]\}.
\]
The corresponding null hypothesis is
\[
    \mathcal{H}_0^{\mathrm{BLP}}:\beta_2=0.
\]
This null asks whether the ML score is linearly related to treatment-effect variation. Thus, a rejection provides evidence of heterogeneity captured by the score $S(X)$.

Sorted-effect methods use the same general logic but test for heterogeneity across groups rather than through a linear slope. Let $G_1,\ldots,G_K$ denote groups formed by sorting observations according to the ML score $S(X)$, and let
\[
    \delta_k := \operatorname{E}\{\delta(X)\mid X\in G_k\}
\]
denote the average treatment effect in group $k$. The corresponding homogeneity null is
\[
    \mathcal{H}_0^{\mathrm{sorted}}:\delta_1=\cdots=\delta_K.
\]
This null asks whether average treatment effects are the same across the groups induced by the ML score. Both approaches, BLP and sorted-effects methods,  therefore test whether heterogeneity is detectable through a particular ML-generated summary, a linear projection in the BLP case or ranked groups in the sorted-effect case. The objective of our test is more basic. We ask whether the data provide evidence that systematic heterogeneity with respect to $X_c$ is present at all. The test therefore does not require heterogeneity to be captured by a linear projection or by a small number of ranked groups.

The null hypothesis considered here also differs from sharp hypotheses of identical unit-level effects studied in randomization-based tests, such as
\[
    \mathcal{H}_0^{\mathrm{sharp}}:
   \operatorname{Var}\{Y(1)-Y(0)\}=0
\]
\citep[e.g.,][]{heck97,ding2016randomization,ding2019decomposing}. Those hypotheses ask whether every individual treatment effect is the same. They can therefore be violated even when treatment-effect variation is idiosyncratic and unrelated to observable characteristics. By contrast, our null is weaker and more targeted. It asks whether the average treatment effect varies in a way that is explainable by the observed covariates $X_c$. This distinction is important in empirical work, where researchers are often interested in whether effects differ across observable and policy-relevant groups rather than whether all individuals have exactly the same effect.

Thus, the proposed test addresses a simple question: do treatment effects vary systematically with $X_c$? The test is omnibus in the sense that it is designed to detect many forms of heterogeneity, including nonlinearities, threshold effects, and interactions among the components of $X_c$. At the same time, it does not require researchers to estimate or interpret the full heterogeneity function. The test can therefore serve as a first step before more detailed analyses of which covariates drive heterogeneity and which groups are most affected.

\section{Testing Heterogeneity under Unconfoundedness}\label{sec: framework}

Consider the potential-outcomes framework introduced in Section
\ref{sec:intuition}. We now formalize the proposed test for treatment-effect heterogeneity under unconfoundedness. Our objective is to construct a test that remains valid when treatment effects are identified using a potentially high-dimensional set of controls and auxiliary quantities are estimated using flexible ML methods.

The development of the test proceeds in four steps. First, we rewrite the null hypothesis of homogeneous treatment effects as a conditional moment restriction involving observable quantities. Second, we transform this conditional restriction into an equivalent collection of unconditional moment conditions that can be tested in practice. Third, we construct locally robust moments that remain valid when the required auxiliary quantities are estimated using ML methods. Fourth, we combine these moments into a test statistic and develop a computationally efficient bootstrap procedure for inference. The resulting procedure provides an omnibus test for treatment-effect heterogeneity that is flexible, computationally tractable, and compatible with a broad class of modern ML estimators.

\subsection{Reformulating the Null Hypothesis}

Throughout this section, we use the notation introduced in Section \ref{sec:intuition}. The observed outcome for unit $i$ is $Y_i=D_iY_i(1)+(1-D_i)Y_i(0)$, and we observe a random sample $\{(Y_i,D_i,X_i)\}_{i=1}^n$, where $n$ denotes the sample size. We first state the identifying assumptions and then rewrite the null hypothesis in a form that can be tested using observed data.

\begin{assumption}
\[
(Y(1),Y(0))\perp\!\!\!\perp D\mid X.
\]
\label{as: unconfoundedness ate}
\end{assumption}

Assumption \ref{as: unconfoundedness ate} requires that, after conditioning on the observed covariates $X$, treatment assignment is independent of the potential outcomes. For this to be the case, $X$ must contain all variables that jointly influence treatment assignment and potential outcomes.

Let
\begin{equation*}
m^D(X):=\operatorname{E}\{D\mid X\}
\end{equation*}
denote the propensity score, that is, the conditional probability of treatment given the observed covariates.

\begin{assumption}\label{as: boundedness of the propensity score ate}
There exists $\epsilon>0$ such that
\[
\epsilon < m^D(X)< 1-\epsilon.
\]
\end{assumption}

Assumption \ref{as: boundedness of the propensity score ate} is a common-support condition. It requires that, for every value of $X$, both treated and untreated units occur with positive probability. This ensures that meaningful comparisons between treated and untreated units can be made.

The null hypothesis in \eqref{eq: null hypothesis} is formulated in terms of potential outcomes, which are only partially observed. We therefore derive an equivalent representation based on quantities that can be estimated from the data. Define
\begin{equation*}
\mu_d(X):=\operatorname{E}\{Y\mid X,D=d\},
\qquad d=0,1.
\end{equation*}
Under Assumption \ref{as: unconfoundedness ate}, these conditional outcome regressions identify the corresponding conditional potential outcomes:
\begin{equation*}
\mu_1(X)=\operatorname{E}\{Y(1)\mid X\} \mbox{ and }
\mu_0(X)=\operatorname{E}\{Y(0)\mid X\}.
\end{equation*}
Hence, the CATE is identified as
\begin{equation*}
\delta(X)=\operatorname{E}\{Y(1)-Y(0)\mid X\}
=
\mu_1(X)-\mu_0(X),
\end{equation*}
and the ATE is identified as
\begin{equation*}
\delta_0
=
\operatorname{E}\{\mu_1(X)-\mu_0(X)\}.
\end{equation*}

Let $X_c$ be a subvector of $X$, with $X_c\in\mathbb{R}^c$. The full covariate vector $X$ is used for identification, whereas $X_c$ contains the covariates with respect to which treatment-effect heterogeneity is evaluated. Applying the Law of Iterated Expectations yields
\begin{equation*}
\operatorname{E}\{Y(1)-Y(0)\mid X_c\}
=
\operatorname{E}\{\mu_1(X)-\mu_0(X)\mid X_c\}.
\end{equation*}
Consequently, the null hypothesis in \eqref{eq: null hypothesis} is equivalent to
\begin{equation}\label{eq: null hypothesis expressed in conditional moment and identified variables}
\mathcal{H}_0:
\operatorname{E}\{\mu_1(X)-\mu_0(X)-\delta_0\mid X_c\}=0.
\end{equation}

Equation \eqref{eq: null hypothesis expressed in conditional moment and identified variables} states that, after removing the average treatment effect $\delta_0$, the remaining treatment-effect variation is conditionally mean independent of $X_c$. This conditional moment restriction is the starting point for the test.

Directly testing \eqref{eq: null hypothesis expressed in conditional moment and identified variables} is difficult because it involves a conditional expectation given $X_c$. We therefore use a strategy from the conditional moment literature and replace the conditional restriction by a collection of unconditional moment restrictions. Intuitively, if treatment effects do not vary systematically with $X_c$, then $\{\mu_1(X)-\mu_0(X)-\delta_0\}$ should be uncorrelated with any function of $X_c$.

Assume, without loss of generality, that $X_c$ has bounded support. Then, by \citet[Theorem 2.2]{bierens2016econometric}, the null hypothesis \eqref{eq: null hypothesis expressed in conditional moment and identified variables} is equivalent to
\begin{equation}\label{eq: non robust hypothesis}
\mathcal{H}_0:
\operatorname{E}
\left\{
[\mu_1(X)-\mu_0(X)-\delta_0]\varphi(t^TX_c)
\right\}
=0
\quad
\forall t\in\mathcal{T},
\end{equation}
where $\varphi$ is an analytic non-polynomial function satisfying
$\partial^l\varphi(0)\neq0$ for all $l\in\mathbb{N}$, and $\mathcal{T}$ is a subset of
$\mathbb{R}^{\dim(X_c)}$ containing a neighborhood of the origin. See Assumption
\ref{as: iid , finite second moments, phi, mu ate} for the precise conditions.

To understand this equivalence, note that
\eqref{eq: null hypothesis expressed in conditional moment and identified variables}
is equivalent to
\begin{equation*}
\operatorname{E}
\left\{
[\mu_1(X)-\mu_0(X)-\delta_0]f(X_c)
\right\}
=0
\end{equation*}
for every square-integrable function $f$. Verifying this condition directly is infeasible because it requires considering an infinite collection of arbitrary functions. The result of \citet{bierens2016econometric} shows that it is sufficient to focus on the structured family, $\{\varphi(t^T\cdot):t\in\mathcal{T}\}$. This reduction is useful because it leads to a test statistic with a simple closed-form representation.

The moment conditions in \eqref{eq: non robust hypothesis} provide a testable characterization of the null hypothesis. A direct implementation, however, would require estimating the unknown functions $\mu_0$ and $\mu_1$. If estimation errors from these first-stage quantities enter the test statistic too strongly, standard inference can fail. To address this issue, we replace the treatment-effect residual in \eqref{eq: non robust hypothesis} by a locally robust score. This construction also uses the propensity score $m^D$.

Consider the doubly robust score
\begin{align*}
V
&:=
\mu_1(X)-\mu_0(X)
+
[Y-\mu_1(X)]\frac{D}{m^D(X)}
-
[Y-\mu_0(X)]\frac{1-D}{1-m^D(X)},
\\
U
&:=
V-\delta_0.
\end{align*}
The quantity $V$ combines outcome-regression adjustment through $\mu_0$ and $\mu_1$ with propensity-score adjustment through $m^D$. Such doubly robust scores are widely used in modern causal inference because they are stable under flexible first-stage estimation \citetext{see, for example, \citealp{chernozhukov2018double}}.

The score $U$ is constructed so that its conditional expectation recovers the treatment-effect residual appearing in \eqref{eq: non robust hypothesis}. Specifically,
\begin{equation*}
\operatorname{E}\{U\mid X\}
=
\mu_1(X)-\mu_0(X)-\delta_0.
\end{equation*}
This follows from
\begin{equation*}
\operatorname{E}\{YD\mid X\}=\mu_1(X)m^D(X) \mbox{ and } \operatorname{E}\{(1-D)Y\mid X\}=\mu_0(X)\{1-m^D(X)\}.
\end{equation*}
Thus, $U$ captures deviations of the conditional treatment effect from the average treatment effect. Under the null hypothesis of homogeneous treatment effects, these deviations should not vary systematically with $X_c$.

Combining this result with \eqref{eq: non robust hypothesis} and the Law of Iterated Expectations yields
\begin{align}
\label{eq: H0 equivalence}
\mathcal{H}_0
\Leftrightarrow
\operatorname{E}\{U\varphi(t^TX_c)\}
=
0
\quad
\forall t\in\mathcal{T}.
\end{align}
Equation \eqref{eq: H0 equivalence} provides the locally robust moment representation used below. Unlike the direct moments in \eqref{eq: non robust hypothesis}, the moments based on $U$ are locally robust with respect to $\mu_0$, $\mu_1$, and $m^D$. Intuitively, small errors in estimating these functions have only a negligible effect on the resulting moments, which makes the construction well suited for flexible ML-based estimation.

\subsection{Locally Robust Test Statistic}\label{sec: test statistic}

Let us now construct a feasible test statistic based on the robust moments in \eqref{eq: H0 equivalence}. If the score $U$ were observed, testing $\mathcal{H}_0$ would be straightforward. In particular, we could measure the extent to which the sample moments
\[
\operatorname{E}_n\{U_i\varphi(t^TX_{c,i})\}
\]
deviate from zero across values of $t$, where $\operatorname{E}_n$ denotes the empirical mean operator, 
\[
\operatorname{E}_ng(\zeta_i)
:=
\frac{1}{n}
\sum_{i=1}^n
g(\zeta_i),
\]
$\zeta_i:=(Y_i,D_i,X_i)$. Following the integrated conditional moment literature, a natural test statistic is
\begin{equation*}
\int
\left|
\sqrt{n}\operatorname{E}_n
\{U_i\varphi(t^TX_{c,i})\}
\right|^2
d\rho(t),
\end{equation*}
where $\rho$ is a probability measure supported on $\mathcal T$.

The challenge is that $U$ depends on the unknown functions $\mu_0$, $\mu_1$, and $m^D$, which must be estimated from the data. To estimate these quantities while avoiding overfitting, we employ cross-fitting. This is particularly important when the nuisance functions are estimated using flexible ML methods. The basic idea is that each observation is evaluated using models that were estimated without using that observation. This reduces overfitting and helps ensure valid inference.

Formally, let $\{\zeta_i\}_{i=1}^n$ denote the observed sample and partition it into $N$ mutually exclusive folds $\{K_1,\ldots,K_N\}$.\footnote{Each observation $\zeta_i$ will belong to a unique fold and the union of such folds gives the entire sample. Formally, $\cup_{s=1}^N K_s=\{\zeta_i\}_{s=1}^n=:K$ and $K_s\cap K_j=\emptyset$ for $s\neq j$.} Let $K(i)$ denote the fold containing observation $i$, and let $K(-i):=K - K(i)$ denote the remaining observations. We estimate the nuisance functions using only the observations in $K(-i)$. Specifically, let $\widehat m^D_{K(-i)}$, $\widehat\mu_{0,K(-i)}$, and $\widehat\mu_{1,K(-i)}$ denote the corresponding estimators of $m^D$, $\mu_0$, and $\mu_1$.

Using these estimators, we construct
\begin{align}
\label{eq: definition of Uhat in ate}
\widehat U_i
:=
\widehat V_i-\widehat\delta,
\end{align}
where
\begin{align*}
\widehat V_i
:=&
\widehat\mu_{1,K(-i)}(X_i)
-
\widehat\mu_{0,K(-i)}(X_i)
\\
&
+
[Y_i-\widehat\mu_{1,K(-i)}(X_i)]
\frac{D_i}
{\widehat m^D_{K(-i)}(X_i)}
\\
&
-
[Y_i-\widehat\mu_{0,K(-i)}(X_i)]
\frac{1-D_i}
{1-\widehat m^D_{K(-i)}(X_i)},
\\
\widehat\delta
:=&
\operatorname{E}_n\{\widehat V_i\}.
\end{align*}

Note that the estimator $\widehat\delta$ is not simply the sample analogue of
\[
\delta_0=\operatorname{E}\{\mu_1(X)-\mu_0(X)\}.
\]
A natural plug-in estimator would replace $\mu_0$ and $\mu_1$ with estimated versions and then average the resulting treatment effects. However, when these quantities are estimated using flexible ML methods, estimation errors can propagate into later stages of the analysis and invalidate standard inference procedures. We therefore use the estimator $\widehat\delta$ because it is locally robust to small errors in the estimation of $\mu_0$, $\mu_1$, and $m^D$; see \citet{chernozhukov2018double}. This robustness property plays a key role in ensuring valid inference while retaining the flexibility of ML estimation.

Our feasible test statistic is
\begin{equation*}
    S_n=
    \int
    \left|
    \sqrt{n}\operatorname{E}_n
    \widehat U_i
    \phi_t(X_{c,i})
    \right|^2
    d\rho(t),
\end{equation*}
where, for notational convenience, we define
$\phi_t(\cdot):=\varphi(t^T\cdot)$.

Intuitively, the statistic measures the extent to which the estimated treatment-effect residuals $\widehat U_i$ are systematically related to functions of the heterogeneity covariates $X_c$. Under the null hypothesis, these empirical moments should be close to zero for all values of $t$. Large values of $S_n$ therefore provide evidence against treatment-effect homogeneity.

Under suitable choices of $\rho$ and $\phi_t$, the statistic admits a simple closed-form representation. In particular, suppose that $\rho$ has a symmetric Fourier transform, denoted by $\mathcal{F}_\rho$, and let
\[
\phi_t(\cdot)=\exp(\bm{i}t^T\cdot),
\]
where $\bm{i}$ denotes the imaginary unit. Then
\begin{equation}
\label{eq: computation ofSn}
S_n
=
\frac{1}{n}
\sum_{i,j}
\widehat U_i
\widehat U_j
\mathcal{F}_\rho(X_{c,i}-X_{c,j})
=
\frac{1}{n}
\bm{\widehat U}^T
\bm{F_\rho}
\bm{\widehat U},
\end{equation}
where
$\bm{\widehat U}
=
(\widehat U_1,\ldots,\widehat U_n)^T$
and $\bm{F_\rho}$ denotes the $n\times n$ matrix with generic element
$\mathcal{F}_\rho(X_{c,i}-X_{c,j})$. Equation \eqref{eq: computation ofSn} is particularly attractive from a computational perspective. The statistic depends only on the estimated quantities $\widehat\mu_0$, $\widehat\mu_1$, and $\widehat m^D$, and does not require any nonparametric smoothing with respect to $X_c$.

We now introduce the regularity conditions required for the asymptotic analysis of the test statistic.

\begin{assumption}\label{as: iid , finite second moments, phi, mu ate}~\\(i) $\{Y_i,D_i,X_i\}_{i=1}^n$ is an iid sample;\\ (ii) $Y$, $\mu_0$, and $\mu_1$ are bounded;\\ (iii) $\mathcal{T}$ is a compact and convex subset of $\mathbb{R}^{\dim(X_c)}$ containing a neighborhood of the origin and $X_c$ is bounded;\\
(iv) $\varphi$ is an analytic non-polynomial function with $\partial^l \varphi(0)\neq 0$ for all $l\in\mathbb{N}$.
\end{assumption}

Parts (i) and (ii) of Assumption \ref{as: iid , finite second moments, phi, mu ate} are standard regularity conditions. Parts (iii) and (iv) ensure that the continuum of unconditional moments in \eqref{eq: H0 equivalence} provides a complete characterization of the original conditional moment restriction. In particular, they guarantee the equivalence between the null hypothesis and the integrated moment conditions used by the test; see \citet[Theorem 2.2]{bierens2016econometric}. The bounded-support assumption on $X_c$ is without loss of generality, since any unbounded covariate can be transformed through a bounded one-to-one mapping without changing the underlying conditional moment restriction.

For any function $f(\zeta)$, with $\zeta:=(Y,D,X)$, define the $L^2(P)$ norm as
\[
\|f\|_{L^2(P)}^2
=
\int f(\zeta)^2 dP(\zeta),
\]
where $P$ denotes the probability law of $\zeta$. Let $\widehat \mu_{1,K-K_j}$ denote the estimator of $\mu_1$ computed on the subsample $K-K_j$, that is, on the sample excluding the observations contained in fold $K_j$. The estimators $\widehat \mu_{0,K-K_j}$ and $\widehat m^D_{K-K_j}$ are defined analogously. Finally, let $\mathcal X$ denote the support of $X$.

\begin{assumption}\label{as: convergence rates for ml estimators ate}
~\\(i) For any $j\in\{1,...,N\}$, \begin{align*}&\|\widehat m^D_{K-K_j} -m^D \|_{L^2(P)}+\|\widehat \mu_{0,K-K_j} -\mu_0 \|_{L^2(P)}+\|\widehat \mu_{1,K-K_j} -\mu_1 \|_{L^2(P)}=o_P(n^{-1/4}).\end{align*} 
(ii) There exists a constant $C>0$ such that for any $j\in\{1,...,N\}$ and $d\in\{0,1\}$, 
\begin{align*}\lim_n\P\left(\sup_{x\in \mathcal{X}}|\widehat \mu_{d,K-K_j}(x)|<C\right)=1.\end{align*}
(iii) There exists $\epsilon>0$ such that for any $j\in\{1,...,N\}$,
\begin{align*}\lim_n\P\left(\epsilon< \inf_{x\in \mathcal{X}}\widehat m^D_{K-K_j}(x)\right)=\lim_n\P\left(\sup_{x\in \mathcal{X}}\widehat m^D_{K-K_j}(x)<1-\epsilon\right)=1.\end{align*}  
\end{assumption}

Assumption \ref{as: convergence rates for ml estimators ate} concerns the quality of the estimators used for the nuisance functions $\mu_0$, $\mu_1$, and $m^D$. Part (i) requires these estimators to be sufficiently accurate in large samples. Formally, the estimation error must converge to zero at a rate faster than $n^{-1/4}$ in the $L^2(P)$ norm. Similar conditions are standard in the double ML literature. For example, the required convergence rates have been established for neural networks by \citet{farrell2021deep} and for random forests by \citet{scornet2015consistency}. Part (ii) requires the estimated outcome regressions $\widehat \mu_{d,K-K_j}$ to remain uniformly bounded over the support $\mathcal X$. Part (iii) requires the estimated propensity score $\widehat m^D_{K-K_j}$ to remain bounded away from zero and one. This condition ensures that the inverse-probability weights appearing in \eqref{eq: definition of Uhat in ate} remain well behaved and do not become excessively large.

The following proposition characterizes the large-sample behavior of the test statistic $S_n$. To state the result, let 
\begin{equation*}
    \iota_t:=\operatorname{E}\{\phi_t(X_c)\}\text{ with }t\in \mathcal{T}\, 
\end{equation*}
and recall that $\zeta_i=(Y_i,D_i,X_i)$. 
\begin{tcolorbox}[colback=gray!6!white, colframe=black]
\begin{proposition}\label{prop: ate statistics asymptotics}
    Let Assumptions \ref{as: unconfoundedness ate}, \ref{as: boundedness of the propensity score ate}, \ref{as: iid , finite second moments, phi, mu ate}, and \ref{as: convergence rates for ml estimators ate}  hold.  Then,

    (a) Under $\mathcal{H}_0$, for $f_t(\zeta)=U[\phi_t(X_c)-\iota_t]$, 
   \begin{equation*}
    S_n \overset{d}{\to} \int_{ }|\mathbb{G}(f_t)|^2 d\rho(t)\, ,   
   \end{equation*}
     where $\{\mathbb{G}(f_t):t\in\mathcal{T}\}$ is a zero-mean tight Gaussian process taking values on $\ell^\infty(\mathcal{T})$, the space of uniformly bounded functionals on $\mathcal{T}$, and characterized by the collection of covariances 
     \begin{gather*}
         \{\operatorname{E}f_{t_1}(\zeta) f_{t_2}(\zeta):t_1,t_2\in\mathcal{T}\}
     \end{gather*}
(b) Under $\mathcal{H}_1$, $S_n \overset{P}{\to} \infty$.
\end{proposition}
\end{tcolorbox} 

Part (a) of Proposition \ref{prop: ate statistics asymptotics} establishes the asymptotic distribution of the test statistic under the null hypothesis of homogeneous treatment effects. Part (b) shows that the test statistic diverges under the alternative hypothesis, implying that the proposed test is consistent against fixed alternatives. The proof is provided in the supplementary material.

Although the limiting distribution in Part (a) is well defined, it depends on unknown features of the data-generating process and is therefore difficult to use directly for inference. In the next subsection, we develop a bootstrap procedure that provides a practical way to approximate the distribution of the test statistic and obtain critical values.

\subsection{A Simple Bootstrap Test}\label{sec: bootstrap test}
To obtain the critical values necessary for testing, we bootstrap our test statistic. Let $\{\xi_i\}_{i=1}^n$ be a sequence of iid bootstrap weights independent from the sample data, with $\E\{\xi\}=1$ and $\operatorname{Var}\{\xi\}=1$. We define the bootstrap counterpart of our statistic as
\begin{equation*}
     S_n^{*}=\int |\sqrt{n}\operatorname{E}_n \{[\widehat U^*_i - \widetilde U_i^*]\phi_t(X_{c,i})\}|^2d\rho(t)
\end{equation*}
where 
\begin{align*}
    \widehat U^*_i=&\xi_i [\widehat V_i-\widehat \delta^*]\\
    \widetilde U_i^*=&\widehat V_i - \widetilde \delta^*\\
    \widehat \delta^*=& \operatorname{E}_n \{\xi_i \widehat V_i\}\\
    \widetilde \delta^*=& \widehat \delta \operatorname{E}_n\{\xi_i\}\, .
\end{align*}
When $\rho$ has a symmetric Fourier transform, $\mathcal{F}_\rho$, and $\phi_t(\cdot)=\exp(\bm{i}t^T\cdot)$, the bootstrapped statistic $S_n^*$ has a simple expression given by
\begin{equation}\label{eq: computation of Sn star}
   \widetilde S_n^{*}= \frac{1}{n}(\bm{\widehat U^* - \widetilde U^*)^T\bm{F_\rho}} (\bm{\widehat U^*} - \bm{\widetilde U^*})\,,
\end{equation}
where $\bm{\widehat U^*}=(\widehat U_1^*,\ldots,\widehat U_n^*)^T$ and $\bm{\widetilde U^*}=(\widetilde U_1^*,\ldots,\widetilde U_n^*)^T$. 

In practice, to obtain the critical value necessary for testing, we can apply the following steps:

\begin{tcolorbox}[colback=green!5!white,colframe=green!75!black,title=Bootstrap test for ATE heterogeneity]
\textbf{Inputs}: $\bm{\widehat U}$, $\bm{F_\rho}$, $S_n$. 
\begin{enumerate}
\item \textbf{For} $b=1:B$ 
\begin{enumerate}
    \item Draw an iid sample $(\xi_1,\ldots,\xi_n)$ from the distribution of $\xi$ such that $\E\{\xi\}=1$ and $\operatorname{Var}\{\xi\}=1$
    \item Compute the bootstrapped statistic $S_{n,b}^*$ according to the expression in  \eqref{eq: computation of Sn star} 
\end{enumerate}
\textbf{EndFor}\\
\textbf{Output} $\{S_{n,b}^*:b=1,\ldots,B\}$.

\item Compute the $1-\alpha$ quantile of $\{S_{n,b}^*:b=1,\ldots,B\}$, say $\widehat q_{1-\alpha}$.
\item If $S_n>\widehat q_{1-\alpha}$ reject $\mathcal{H}_0$.
\end{enumerate}
\end{tcolorbox}
The above procedure is based on vector and quadratic forms computations, so it is straightforward to implement. Note that it does not require to compute the ML estimators at each bootstrap iteration. This is computationally convenient. \\
We next establish the validity of this bootstrap procedure. Let us denote with $\Pr^*$ the probability that considers as random only the bootstrap weights $\{\xi_i\}_{i=1}^n$, that is, the probability conditional to the sample data.  We define formally the $1-\alpha$ quantile of the bootstrap distribution of $S_n^*$, $\widehat q_{1-\alpha}$, as 
\begin{equation}\label{eq: bootstrap quantile ate}
    \widehat q_{1-\alpha}=\inf \left\{q: {\Pr}^*(S_n^*\leq q)\geq 1-\alpha\right\}\, .
\end{equation}

\begin{tcolorbox}[colback=gray!6!white, colframe=black]
\begin{proposition}\label{prop: bootstrap validity ate}
    Let Assumptions \ref{as: unconfoundedness ate}, \ref{as: boundedness of the propensity score ate}, \ref{as: iid , finite second moments, phi, mu ate}, and \ref{as: convergence rates for ml estimators ate} hold. Then, \\
    (a) Under $\mathcal{H}_0$, 
    \begin{equation*}
       \Pr(S_n>\widehat q_{1-\alpha})\rightarrow\alpha
    \end{equation*}
    (b) Under $\mathcal{H}_1$, 
    \begin{equation*}
        \Pr(S_n>\widehat q_{1-\alpha})\rightarrow 1\, .
    \end{equation*}
    
\end{proposition}
\end{tcolorbox}
The above result is shown in the supplementary material.
Part (a) of the above proposition shows that our bootstrap test controls the size asymptotically. Part (b) shows that it can detect violations of the null hypothesis.   

\section{Beyond Unconfoundedness}\label{sec: beyond}

Section \ref{sec: framework} develops our testing procedure under identification by unconfoundedness. The same ideas, however, extend naturally to other widely used causal inference frameworks. In this section, we consider Difference-in-Differences (DiD) models with panel and repeated cross section data, as well as Local Average Treatment Effects (LATE) identified through instrumental variables.

The construction closely parallels that developed in Section \ref{sec: framework}. In each case, we reformulate the null hypothesis of homogeneous treatment effects in terms of observable quantities, construct locally robust moment conditions, and build a test statistic based on cross-fitted estimators of the required nuisance functions. The resulting procedures retain the same ability to accommodate flexible ML estimators while maintaining valid inference.

To avoid repetition, we focus on the framework-specific assumptions and formal results. Detailed derivations of the moment conditions, test statistics, asymptotic distributions, and bootstrap procedures are provided in the Appendix.

\subsection{Difference-in-Differences with Panel Data}\label{sec: did}

We first consider a Difference-in-Differences (DiD) design with two periods, $t=0,1$, and panel data. Let $G_i\in\{0,1\}$ indicate whether unit $i$ belongs to the treated group. Treated units receive treatment only in period $1$, while control units are never treated. Let $Y_{i,t}(d)$ denote the potential outcome in period $t$ under treatment status $d\in\{0,1\}$, and let $Y_{i,t}$ denote the observed outcome. We observe
\begin{equation*}
\{Y_{i,0},Y_{i,1},G_i,X_i\}_{i=1}^n,
\end{equation*}
where $X_i$ is a vector of pre-treatment covariates. Define
\begin{equation*}
\Delta Y_i:=Y_{i,1}-Y_{i,0},
\qquad
\pi_0:=\operatorname{E}\{G_i\},
\qquad
m^G(X):=\operatorname{E}\{G\mid X\}.
\end{equation*}

Our objective is to test whether the average treatment effect on the treated varies systematically with the heterogeneity covariates $X_c\subseteq X$. Formally, we consider
\begin{gather}
\label{eq: null hypothesis did unobservable}
\mathcal{H}_0^{DiD}:
\operatorname{E}\{Y_1(1)-Y_1(0)\mid G=1,X_c\}
=
\delta_0
\text{ for some }\delta_0\in\mathbb{R},
\\
\text{versus }\mathcal{H}_1^{DiD}:\mathcal{H}_0^{DiD,c}\,,
\nonumber
\end{gather}
where $\mathcal{H}_0^{DiD,c}$ denotes the logical complement of $\mathcal{H}_0^{DiD}$.

Identification relies on the standard conditional parallel-trends assumption.

\begin{assumption}\label{as: coditional common trend in did}
\begin{equation*}
\operatorname{E}\{Y_{1}(0)-Y_{0}(0)\mid G=1,X\}
=
\operatorname{E}\{Y_{1}(0)-Y_{0}(0)\mid G=0,X\}.
\end{equation*}
\end{assumption}

\begin{assumption}\label{as: support of the treated in did}
(i) $\pi_0>0$; \qquad
(ii) there exists $\epsilon\in(0,1)$ such that
$m^G(X)<1-\epsilon$.
\end{assumption}

\begin{assumption}\label{as: propensoity score did}
\begin{equation*}
\operatorname{E}\{G\mid X_c\}>0.
\end{equation*}
\end{assumption}

Assumption \ref{as: coditional common trend in did} requires that, after conditioning on $X$, treated and control units would have experienced the same average outcome trend in the absence of treatment. Assumption \ref{as: support of the treated in did} ensures that treated units exist and that suitable control units are available across the relevant support of $X$. Assumption \ref{as: propensoity score did} ensures that the conditional treatment effect for the treated is well defined for the values of $X_c$ considered in the test.

Define
\begin{equation*}
\mu^{Y_1-Y_0}_{0}(X)
:=
\operatorname{E}\{\Delta Y\mid G=0,X\}.
\end{equation*}
Following the same logic as in Section \ref{sec: framework}, the null hypothesis can be written as a collection of locally robust moment restrictions. Let
\begin{gather}
V^{DiD}
:=
\left[\Delta Y-\mu^{Y_1-Y_0}_{0}(X)\right]
\frac{G-m^G(X)}
{1-m^G(X)},
\nonumber\\
U^{DiD}
:=
V^{DiD}-\delta_0G.
\label{eq: definition of Udid and Vdid}
\end{gather}
Then
\begin{equation}\label{eq: equivalent null hypothesis in DiD}
\mathcal{H}_0^{DiD}
\Leftrightarrow
\operatorname{E}\{U^{DiD}\varphi(t^TX_c)\}
=
0
\qquad
\forall t\in\mathcal{T}.
\end{equation}
A detailed derivation of Equation \eqref{eq: equivalent null hypothesis in DiD} is provided in Appendix \ref{sec: proof of H0DiD and H0DiDrcs}.
As in Section \ref{sec: test statistic}, we estimate the nuisance functions using cross-fitting. Let $\widehat \mu^{Y_1-Y_0}_{0,K(-i)}$ and $\widehat m^G_{K(-i)}$ denote estimators obtained without using the fold containing observation $i$. Define
\begin{gather*}
\widehat V^{DiD}_i
:=
\left[\Delta Y_i-\widehat \mu^{Y_1-Y_0}_{0,K(-i)}(X_i)\right]
\frac{G_i-\widehat m^G_{K(-i)}(X_i)}
{1-\widehat m^G_{K(-i)}(X_i)},
\\
\widehat \pi
:=
\operatorname{E}_n\{G_i\},
\qquad
\widehat \delta^{DiD}
:=
\frac{\operatorname{E}_n\{\widehat V_i^{DiD}\}}
{\widehat \pi},
\\
\widehat U_i^{DiD}
:=
\widehat V_i^{DiD}
-
\widehat \delta^{DiD}G_i.
\end{gather*}

The resulting test statistic is
\begin{equation}
\label{test statistic for did}
S_n^{DiD}
=
\int
\left|
\sqrt{n}\operatorname{E}_n
\{\widehat U_i^{DiD}\phi_t(X_{c,i})\}
\right|^2
d\rho(t),
\end{equation}
where $\phi_t(X_c)=\varphi(t^TX_c)$. As in Section \ref{sec: test statistic}, choosing
\begin{equation*}
\phi_t(\cdot)=\exp(\bm{i}t^T\cdot)
\end{equation*}
yields the computationally convenient representation
\begin{equation*}
S_n^{DiD}
=
\frac{1}{n}
(\bm{\widehat U}^{DiD})^T
\bm{F_\rho}
\bm{\widehat U}^{DiD},
\end{equation*}
where
\begin{equation*}
\bm{\widehat U}^{DiD}
=
(\widehat U_1^{DiD},\ldots,\widehat U_n^{DiD})^T
\end{equation*}
and  $\bm{F_\rho}$ is defined in Section \ref{sec: test statistic}.
We impose the following regularity conditions. Let $\zeta:=(Y_0,Y_1,G,X)$, let $P$ denote the distribution of $\zeta$, and let $\mathcal X$ denote the support of $X$. The norm $\|\cdot\|_{L^2(P)}$ is defined as in Section \ref{sec: test statistic}.

\begin{assumption}\label{as: iid , finite second moments, phi, mu did}
~\\
(i) $\{Y_{0,i},Y_{1,i},G_i,X_i\}_{i=1}^n$ is an iid sample;
\\
(ii) $Y_1-Y_0$ and $\mu_{0}^{Y_1-Y_0}$ are bounded;
\\
(iii) $\mathcal{T}$ is a compact subset of $\mathbb{R}^{\dim(X_c)}$ containing a neighborhood of the origin and $X_c$ is bounded;
\\
(iv) $\varphi$ is an analytic non-polynomial function with $\partial^l\varphi(0)\neq 0$ for all $l\in\mathbb{N}$.
\end{assumption}

\begin{assumption}\label{as: convergence rates for ml estimators did}
~\\
(i) For any $j\in\{1,\ldots,N\}$,
\begin{equation*}
\|
\widehat m^G_{K-K_j}(X)-m^G(X)
\|_{L^2(P)}
+
\|
\widehat \mu^{Y_1-Y_0}_{0,K-K_j}(X)
-
\mu^{Y_1-Y_0}_{0}(X)
\|_{L^2(P)}
=
o_P(n^{-1/4}).
\end{equation*}

(ii) There exists a constant $C>0$ such that for any $j\in\{1,\ldots,N\}$,
\begin{equation*}
\lim_n
\P\left(
\sup_{x\in\mathcal X}
|\widehat\mu^{Y_1-Y_0}_{0,K-K_j}(x)|+ \sup_{x\in\mathcal{X}}|\widehat m^G_{K-K_j}(x)|
<C
\right)
=
1.
\end{equation*}

(iii) There exists $\epsilon>0$ such that for any $j\in\{1,\ldots,N\}$,
\begin{equation*}
\lim_n
\P\left(
\sup_{x\in\mathcal X}
\widehat m^G_{K-K_j}(x)
<
1-\epsilon
\right)
=
1.
\end{equation*}
\end{assumption}

Assumptions \ref{as: iid , finite second moments, phi, mu did} and \ref{as: convergence rates for ml estimators did} are the DiD analogues of the regularity conditions imposed in the unconfoundedness setting. They ensure that the integrated moment representation characterizes the null hypothesis and that the nuisance estimators are sufficiently accurate for locally robust inference. As before, the required convergence rates can be satisfied by a broad class of modern ML estimators.

\begin{tcolorbox}[colback=gray!6!white, colframe=black]
\begin{proposition}\label{prop: did statistics asymptotics}
Let Assumptions \ref{as: coditional common trend in did}, \ref{as: support of the treated in did}, \ref{as: propensoity score did}, \ref{as: iid , finite second moments, phi, mu did}, and \ref{as: convergence rates for ml estimators did} hold. Then:

(a) For
\begin{equation*}
f_t(\zeta)
:=
U^{DiD}\phi_t(X_c)
-
[\operatorname{E}\{G\phi_t(X_c)\}]
\left[
\frac{V^{DiD}}{\pi_0}
-
\delta_0
-
\frac{\delta_0}{\pi_0}(G-\pi_0)
\right],
\end{equation*}
under $\mathcal{H}_0^{DiD}$,
\begin{equation*}
S_n^{DiD}
\overset{d}{\to}
\int
|\mathbb{G}(f_t)|^2
d\rho(t),
\end{equation*}
where $\{\mathbb{G}(f_t):t\in\mathcal{T}\}$ is a zero-mean tight Gaussian process taking values on $\ell^\infty(\mathcal{T})$ and characterized by the collection of covariances
\begin{equation*}
\{
\operatorname{E}f_{t_1}(\zeta)f_{t_2}(\zeta)
:
t_1,t_2\in\mathcal{T}
\}.
\end{equation*}

(b) Under $\mathcal{H}_1^{DiD}$,
\begin{equation*}
S_n^{DiD}
\overset{P}{\to}
\infty.
\end{equation*}
\end{proposition}
\end{tcolorbox}

Part (a) establishes the asymptotic distribution of the test statistic under the null hypothesis of homogeneous average treatment effects on the treated. Part (b) shows that the statistic diverges under the alternative, implying that the test is consistent against fixed alternatives.  The proof is provided in the supplementary material.

We obtain critical values using a bootstrap procedure analogous to that in Section \ref{sec: bootstrap test}. The nuisance functions are estimated only once and are kept fixed across bootstrap iterations. The bootstrap recomputes the low-dimensional quantities that enter the DiD score, namely the treated share and the ATT estimate. Let $\{\xi_i\}_{i=1}^n$ be an iid sequence of bootstrap weights independent from the sample data, with $\operatorname{E}\{\xi\}=\operatorname{Var}\{\xi\}=1$. We define
\begin{gather*}
    \widehat \pi^*:=\operatorname{E}_n\{\xi_i G_i\}\,,\quad
    \widehat \delta^*:= \operatorname{E}_n \{\xi_i \widehat V_i\}/\widehat \pi^*\,,\quad
    \widehat U_i^{DiD*}:=\xi_i [\widehat V_i - \widehat \delta^* G_i]\, .
\end{gather*}
Then, the bootstrap counterpart of the test statistic $S_n^{DiD}$ is 
\begin{equation}
    S_n^{DiD *}=\int_{ }\left|\sqrt{n}\operatorname{E}_n \{ [\widehat U^{DiD*}_i - \widehat U^{DiD}_i ]\phi_t(X_{c,i})\}\right|^2 d\rho(t)\,,
\end{equation}
By setting $\phi_t(\cdot)=\exp(\textbf{i} t^T\cdot)$, the bootstrapped statistic $S_n^{DID *}$ can be computed similarly to $S_n^{DiD}$, as follows
\begin{equation*}
    S_n^{DiD*}=\frac{1}{n}(\bm{\widehat U^{DiD*}}-\bm{\widehat U^{DiD}} )^T\bm{F_\rho} (\bm{\widehat U^{DiD*}-\widehat U^{DiD}})\,,
\end{equation*}
where $\bm{\widehat U^{DiD*}}=(\widehat U^{DiD*}_1,\ldots,\widehat U^{DiD*}_n)^T$ and $\bm{\widehat U^{DiD}}=(\widehat U^{DiD}_1,\ldots,\widehat U^{DiD}_n)^T$. The bootstrap test can be implemented by the Monte Carlo procedure described in Section \ref{sec: bootstrap test}.

Let $\widehat q_{1-\alpha}^{DiD}$ denote the empirical $(1-\alpha)$ quantile of the bootstrap statistics generated by the procedure above. The following proposition establishes the validity of this design-specific bootstrap critical value.

\begin{tcolorbox}[colback=gray!6!white, colframe=black]
\begin{proposition}\label{prop: bootstrap validity did}
Let Assumptions \ref{as: coditional common trend in did}, \ref{as: support of the treated in did}, \ref{as: propensoity score did}, \ref{as: iid , finite second moments, phi, mu did}, and \ref{as: convergence rates for ml estimators did} hold. Then:

(a) Under $\mathcal{H}_0^{DiD}$,
\begin{equation*}
\Pr(S_n^{DiD}>\widehat q_{1-\alpha}^{DiD})
\rightarrow
\alpha.
\end{equation*}

(b) Under $\mathcal{H}_1^{DiD}$,
\begin{equation*}
\Pr(S_n^{DiD}>\widehat q_{1-\alpha}^{DiD})
\rightarrow
1.
\end{equation*}
\end{proposition}
\end{tcolorbox}

Proposition \ref{prop: bootstrap validity did} establishes that the bootstrap test has asymptotically correct size under the null hypothesis and is consistent under fixed alternatives. The proof is provided in the supplementary material.

\subsection{Difference-in-Differences with Repeated Cross Sections}
\label{sec: did repeated cross sections}

We next consider a Difference-in-Differences design with repeated cross sections. In contrast to the panel-data setting, the same unit is not observed before and after treatment. Instead, the data contain different units observed either in the pre-treatment period or in the post-treatment period. Let $T_i\in\{0,1\}$ indicate whether unit $i$ is observed in the post-treatment period, and let $G_i\in\{0,1\}$ indicate whether unit $i$ belongs to the treated group. We observe
\begin{equation*}
\{Y_i,G_i,T_i,X_i\}_{i=1}^n,
\end{equation*}
where $Y_i$ is the observed outcome and $X_i$ is a vector of pre-treatment covariates. We maintain the conditional parallel-trends and support conditions from Assumptions \ref{as: coditional common trend in did}, \ref{as: support of the treated in did}, and \ref{as: propensoity score did}. As before, let
\begin{equation*}
m^G(X):=\operatorname{E}\{G\mid X\},
\qquad
\pi_0:=\operatorname{E}\{G\}.
\end{equation*}

The repeated-cross-section sampling scheme is formalized as follows.

\begin{assumption}\label{as: mixture for repeated cross section}
The observed sample is iid from
\begin{align*}
\Pr(Y=y,G=g,X=x,T=t)
=&
\lambda\cdot t\cdot {\Pr}_{1}(Y_1=y,G=g,X=x)
\\
&+
(1-\lambda)\cdot(1-t)\cdot {\Pr}_{0}(Y_0=y,G=g,X=x),
\end{align*}
where $\lambda\in(0,1)$, $\Pr_1$ is the distribution of $(Y_1,G,X)$ conditional on $T=1$, and $\Pr_0$ is the distribution of $(Y_0,G,X)$ conditional on $T=0$.
\end{assumption}

Assumption \ref{as: mixture for repeated cross section} states that the observed data combine two cross sections: one from the pre-treatment period and one from the post-treatment period. In particular, $\lambda=\Pr(T=1)$ denotes the population share of observations from the post-treatment period.

Our objective is again to test whether the treatment effect on the treated varies systematically with $X_c$. Formally, we consider
\begin{gather}
\label{H0 did rcs}
\mathcal{H}_0^{DiDrcs}:
\operatorname{E}\{Y_{1}(1)-Y_{1}(0)\mid G=1,X_c\}
=
\delta_0
\text{ for some }\delta_0\in\mathbb{R},
\\
\text{versus }\mathcal{H}_1^{DiDrcs}:\mathcal{H}_0^{DiDrcs,c},
\nonumber
\end{gather}
where $\mathcal{H}_0^{DiDrcs,c}$ denotes the logical complement of $\mathcal{H}_0^{DiDrcs}$.

Following the same logic as in Section \ref{sec: did}, the null hypothesis can be represented through locally robust moments. Define
\begin{equation*}
\mu_0^Y(X):=\operatorname{E}\{Y\mid X,G=0\},
\qquad
\mu_0^{YT}(X):=\operatorname{E}\{YT\mid X,G=0\}.
\end{equation*}
Let
\begin{align}
W(\lambda)
:=&
\left[
\frac{YT-\mu_0^{YT}(X)}
{\lambda(1-\lambda)}
-
\frac{Y-\mu_0^Y(X)}
{1-\lambda}
\right]
\frac{G-m^G(X)}
{1-m^G(X)},
\nonumber\\
U^{DiDrcs}
:=&
W(\lambda)-\delta_0G.
\label{eq: definition of U did rcs}
\end{align}
Then
\begin{equation}\label{eq: equivalent null hypothesis in DiD rcs}
\mathcal{H}_0^{DiDrcs}
\Leftrightarrow
\operatorname{E}\{U^{DiDrcs}\varphi(t^TX_c)\}
=
0
\qquad
\forall t\in\mathcal{T}.
\end{equation}

A detailed derivation of Equation \eqref{eq: equivalent null hypothesis in DiD rcs} is provided in Appendix \ref{sec: proof of H0DiD and H0DiDrcs}.
The score $U^{DiDrcs}$ plays the same role as $U^{DiD}$ in the panel-data case. It converts the null hypothesis into moments that can be estimated using observed data and remains locally robust with respect to the first-stage nuisance estimates. The additional feature in the repeated-cross-section setting is that the score depends on the sampling share $\lambda$, which also has to be estimated.

We estimate $\mu_0^Y$, $\mu_0^{YT}$, and $m^G$ by cross-fitting. Let $\widehat\mu^Y_{0,K(-i)}$, $\widehat\mu^{YT}_{0,K(-i)}$, and $\widehat m^G_{K(-i)}$ denote estimators obtained without using the fold containing observation $i$. Define
\begin{gather*}
\widehat\lambda
:=
\operatorname{E}_n\{T_i\},
\qquad
\widehat\pi
:=
\operatorname{E}_n\{G_i\},
\\
\widehat W_i(\widehat\lambda)
:=
\left[
\frac{Y_iT_i-\widehat\mu^{YT}_{0,K(-i)}(X_i)}
{\widehat\lambda(1-\widehat\lambda)}
-
\frac{Y_i-\widehat\mu^Y_{0,K(-i)}(X_i)}
{1-\widehat\lambda}
\right]
\frac{G_i-\widehat m^G_{K(-i)}(X_i)}
{1-\widehat m^G_{K(-i)}(X_i)},
\\
\widehat\delta^{DiDrcs}
:=
\frac{\operatorname{E}_n\{\widehat W_i(\widehat\lambda)\}}
{\widehat\pi},
\qquad
\widehat U_i^{DiDrcs}
:=
\widehat W_i(\widehat\lambda)
-
\widehat\delta^{DiDrcs}G_i.
\end{gather*}

The resulting test statistic is
\begin{equation}
\label{eq: test statistic did rcs}
S_n^{DiDrcs}
=
\int
\left|
\sqrt{n}
\operatorname{E}_n
\{\widehat U_i^{DiDrcs}\phi_t(X_{c,i})\}
\right|^2
d\rho(t),
\end{equation}
where $\phi_t(X_c)=\varphi(t^TX_c)$. As before, choosing
\begin{equation*}
\phi_t(\cdot)=\exp(\bm{i}t^T\cdot)
\end{equation*}
gives the closed-form expression
\begin{equation*}
S_n^{DiDrcs}
=
\frac{1}{n}
(\bm{\widehat U}^{DiDrcs})^T
\bm{F_\rho}
\bm{\widehat U}^{DiDrcs},
\end{equation*}
where
\begin{equation*}
\bm{\widehat U}^{DiDrcs}
=
(\widehat U_1^{DiDrcs},\ldots,\widehat U_n^{DiDrcs})^T.
\end{equation*}

We impose the following regularity conditions. Let $\zeta:=(Y,G,T,X)$, let $P$ denote the distribution of $\zeta$, and let $\mathcal X$ denote the support of $X$. The norm $\|\cdot\|_{L^2(P)}$ is defined as in Section \ref{sec: test statistic}.

\begin{assumption}\label{as: iid , finite second moments, phi, mu did rcs}
~\\
(i) $\{Y_i,T_i,G_i,X_i\}_{i=1}^n$ is an iid sample;
\\
(ii) $Y$, $\mu_0^Y$, and $\mu_0^{YT}$ are bounded;
\\
(iii) $\mathcal{T}$ is a compact subset of $\mathbb{R}^c$ containing a neighborhood of the origin and $X_c$ is bounded;
\\
(iv) $\varphi$ is an analytic non-polynomial function with $\partial^l\varphi(0)\neq 0$ for all $l\in\mathbb{N}$.
\end{assumption}

\begin{assumption}\label{as: convergence rates for ml estimators did rcs}
~\\
(i) For any $j\in\{1,\ldots,N\}$,
\begin{align*}
&
\|
\widehat m^G_{K-K_j}(X)-m^G(X)
\|_{L^2(P)}
+
\|
\widehat\mu^Y_{0,K-K_j}(X)
-
\mu^Y_0(X)
\|_{L^2(P)}
\\
&
\qquad
+
\|
\widehat\mu^{YT}_{0,K-K_j}(X)
-
\mu^{YT}_0(X)
\|_{L^2(P)}
=
o_P(n^{-1/4});
\end{align*}

(ii) there exists a constant $C>0$ such that for any $j\in\{1,\ldots,N\}$,
\begin{equation*}
\lim_n
\P\left(
\sup_{x\in\mathcal X}
|\widehat\mu^Y_{0,K-K_j}(x)|
+
\sup_{x\in\mathcal X}
|\widehat\mu^{YT}_{0,K-K_j}(x)| + \sup_{x\in\mathcal{X}}|\widehat m^G_{K-K_j}(x)|
<C
\right)
=
1;
\end{equation*}

(iii) there exists $\epsilon>0$ such that for any $j\in\{1,\ldots,N\}$,
\begin{equation*}
\lim_n
\P\left(
\sup_{x\in\mathcal X}
\widehat m^G_{K-K_j}(x)
<
1-\epsilon
\right)
=
1.
\end{equation*}
\end{assumption}

Assumptions \ref{as: iid , finite second moments, phi, mu did rcs} and \ref{as: convergence rates for ml estimators did rcs} are the repeated-cross-section analogues of the regularity conditions used above. They ensure that the integrated moment representation remains valid, that the nuisance estimators are sufficiently accurate for locally robust inference, and that the inverse-probability weights remain stable.

\begin{tcolorbox}[colback=gray!6!white, colframe=black]
\begin{proposition}\label{prop: did rcs statistics asymptotics}
Let Assumptions \ref{as: coditional common trend in did}, \ref{as: support of the treated in did}, \ref{as: propensoity score did}, \ref{as: mixture for repeated cross section}, \ref{as: iid , finite second moments, phi, mu did rcs}, and \ref{as: convergence rates for ml estimators did rcs} hold. Then:

(a) For
\begin{align*}
f_t(\zeta)
:=&
U^{DiDrcs}\phi_t(X_c)
+
\left[
\operatorname{E}
\left\{
\frac{\partial}{\partial\lambda}
W(\lambda)
\phi_t(X_c)
\right\}
\right]
(T-\lambda)
\\
&-
[\operatorname{E}\{G\phi_t(X_c)\}]
\left[
\frac{W(\lambda)}{\pi_0}
-
\delta_0
-
\frac{\delta_0}{\pi_0}(G-\pi_0)
+
\left(
\operatorname{E}
\left\{
\frac{\partial}{\partial\lambda}
W(\lambda)
\right\}
\right)
\frac{T-\lambda}{\pi_0}
\right],
\end{align*}
under $\mathcal{H}_0^{DiDrcs}$,
\begin{equation*}
S_n^{DiDrcs}
\overset{d}{\to}
\int
|\mathbb{G}(f_t)|^2
d\rho(t),
\end{equation*}
where $\{\mathbb{G}(f_t):t\in\mathcal{T}\}$ is a zero-mean tight Gaussian process taking values on $\ell^\infty(\mathcal{T})$ and characterized by the collection of covariances
\begin{equation*}
\{
\operatorname{E}f_{t_1}(\zeta)f_{t_2}(\zeta)
:
t_1,t_2\in\mathcal{T}
\}.
\end{equation*}

(b) Under $\mathcal{H}_1^{DiDrcs}$,
\begin{equation*}
S_n^{DiDrcs}
\overset{P}{\to}
\infty.
\end{equation*}
\end{proposition}
\end{tcolorbox}

Part (a) establishes the asymptotic distribution of the test statistic under the null hypothesis of homogeneous treatment effects on the treated, accounting for estimation of $\lambda$ and $\pi_0$. Part (b) shows that the statistic diverges under the alternative, implying consistency against fixed alternatives. The proof is provided in the supplementary material.

We obtain critical values using a bootstrap procedure analogous to that in Section \ref{sec: bootstrap test}. The nuisance functions are estimated only once and are kept fixed across bootstrap iterations. Compared with the panel-data case, the bootstrap must also account for the estimation of the post-treatment sampling share $\lambda$, which enters the score $W(\lambda)$. Let $\{\xi_i\}_{i=1}^n$ be a sequence of iid bootstrap weights independent from the sample data, with $\operatorname{E}\{\xi\}=1$ and $\text{Var}\{\xi\}=1$. We define 
\begin{gather*}
    \widehat \lambda^*:=\operatorname{E}_n\{\xi_i\,T_i\}\,,\quad\widehat \pi^*:=\operatorname{E}_n \{\xi_i G_i\}\, ,\quad
    \widehat \delta^*:=\operatorname{E}_n \{\xi_i\widehat W_i(\widehat \lambda ^*)\}/\widehat \pi^*\,,\\
    \widehat U_i^{DiDrcs*}:=\xi_i[\widehat W_i(\widehat \lambda ^*) - \widehat \delta^* G_i]\,,
\end{gather*}
 and 
 \begin{gather*}
     \widetilde \lambda^*:=\operatorname{E}_n\{\xi_i\}\widehat \lambda\,,\quad\widetilde \pi^*:=\operatorname{E}_n \{\xi_i\} \widehat \pi\, ,\quad
     \widetilde \delta^*:=\operatorname{E}_n\{\xi_i\} \operatorname{E}_n \{\widehat W_i(\widetilde \lambda^*)\}/\widetilde \pi^*=\operatorname{E}_n \{\widehat W_i(\widetilde \lambda^*)\}/\widehat \pi\,,\\
     \widetilde U_n^{DiDrcs*}:=\widehat W_i(\widetilde \lambda^*) - \widetilde \delta^* G_i\, .
 \end{gather*}
Then, the bootstrap counterpart of our statistic is 
\begin{equation}
S_n^{DiDrcs*}:=\int_{ }|\sqrt{n}\operatorname{E}_n [\widehat U_i^{DiDrcs*} - \widetilde U^{DiDrcs*}_i] \phi_t(X_{c,i})|^2 d \rho(t)\, .
\end{equation}

By setting $\phi_t(\cdot)=\exp(\textbf{i} t^T\cdot)$, we obtain the following simple expression for the bootstrapped statistic 
\begin{equation*}
S_n^{DiDrcs*}=\frac{1}{n}(\bm{\widehat U ^{DiDrcs*}} - \bm{\widetilde U ^{DiDrcs*}})^T \bm{F_\rho}\;(\bm{\widehat U ^{DiDrcs*}} - \bm{\widetilde U ^{DiDrcs*}})\,,
\end{equation*}
where $[\bm{\widehat U ^{DiDrcs*}}=(\widehat U ^{DiDrcs*}_1,\ldots,\widehat U ^{DiDrcs*}_n)^T$ and $\bm{\widetilde U ^{DiDrcs*}}=(\widetilde U ^{DiDrcs*}_1,\ldots,\widetilde U ^{DiDrcs*}_n)^T$. The bootstrap scheme can then be implemented by the Monte Carlo procedure described in Section \ref{sec: bootstrap test}.  We remark that the above  bootstrap scheme is not  based on a multiplier bootstrap version of the influence function representation of the empirical process at the basis of the test statistic. Such a  version would have a rather complicated expression.  The bootstrap scheme in the present section can be seen as a hybrid between a weighed bootstrap and a multiplier bootstrap.

Let $\widehat q_{1-\alpha}^{DiDrcs}$ denote the empirical $(1-\alpha)$ quantile of the bootstrap statistics generated by the procedure above. The following proposition establishes the validity of this repeated-cross-section bootstrap critical value.

\begin{tcolorbox}[colback=gray!6!white, colframe=black]
\begin{proposition}\label{prop: did rcs statistics boostrap}
Let Assumptions \ref{as: coditional common trend in did}, \ref{as: support of the treated in did}, \ref{as: propensoity score did}, \ref{as: mixture for repeated cross section}, \ref{as: iid , finite second moments, phi, mu did rcs}, and \ref{as: convergence rates for ml estimators did rcs} hold. Then:

(a) Under $\mathcal{H}_0^{DiDrcs}$,
\begin{equation*}
\Pr(S_n^{DiDrcs}>\widehat q_{1-\alpha}^{DiDrcs})
\rightarrow
\alpha.
\end{equation*}

(b) Under $\mathcal{H}_1^{DiDrcs}$,
\begin{equation*}
\Pr(S_n^{DiDrcs}>\widehat q_{1-\alpha}^{DiDrcs})
\rightarrow
1.
\end{equation*}
\end{proposition}
\end{tcolorbox}

Proposition \ref{prop: did rcs statistics boostrap} establishes that the bootstrap test has asymptotically correct size under the null hypothesis and is consistent under fixed alternatives. The proof is provided in the supplementary material.

\subsection{Local Average Treatment Effects}\label{sec: late}

We finally consider settings with a binary instrumental variable. In contrast to the unconfoundedness case, treatment assignment may be endogenous even after conditioning on $X$. Identification instead relies on an instrument $Z\in\{0,1\}$ that affects treatment take-up while satisfying the conditions stated below. Let $Y(d)$ denote the potential outcome under treatment status $d\in\{0,1\}$, and let $D(z)\in\{0,1\}$ denote the potential treatment status under instrument value $z\in\{0,1\}$. We observe an iid sample
\begin{equation*}
\{Y_i,D_i,Z_i,X_i\}_{i=1}^n,
\end{equation*}
where $D_i=D_i(Z_i)$ and
\begin{equation*}
Y_i=D_iY_i(1)+(1-D_i)Y_i(0).
\end{equation*}
Define the instrument propensity score
\begin{equation*}
m^Z(X):=\operatorname{E}\{Z\mid X\}.
\end{equation*}

The target parameter is the Local Average Treatment Effect (LATE), that is, the average treatment effect for compliers, defined as units whose treatment status is shifted by the instrument. Our objective is to test whether this local treatment effect varies systematically with the heterogeneity covariates $X_c\subseteq X$.

\begin{assumption}\label{as: independence in late}
\[
(Y(1),Y(0),D(1),D(0))\perp\!\!\!\perp Z\mid X.
\]
\end{assumption}

\begin{assumption}\label{as: monotonicity in late}
\[
\Pr(D(1)<D(0))=0.
\]
\end{assumption}

\begin{assumption}\label{assumption: conditional relevance late}
\[
\Pr(D(1)>D(0)\mid X_c=u)>0
\quad
\text{for all }u\in \operatorname{Support}(X_c).
\]
\end{assumption}

Assumption \ref{as: independence in late} requires the instrument to be conditionally exogenous given $X$. Assumption \ref{as: monotonicity in late} rules out defiers. Assumption \ref{assumption: conditional relevance late} ensures that compliers exist for each value of $X_c$, so that the conditional LATE is well defined.

Our null hypothesis is
\begin{gather}
\label{eq: null hypothesis in late}
\mathcal{H}_0^{LATE}:
\operatorname{E}\{Y(1)-Y(0)\mid D(1)>D(0),X_c\}
=
\delta_0
\text{ for some fixed constant }\delta_0\in\mathbb{R},
\\
\text{versus }
\mathcal{H}_1^{LATE}:\mathcal{H}_0^{LATE,c},
\nonumber
\end{gather}
where $\mathcal{H}_0^{LATE,c}$ denotes the logical complement of $\mathcal{H}_0^{LATE}$. Under the null hypothesis, the conditional LATE is constant in $X_c$ and coincides with the unconditional LATE, $\delta_0$.

We also impose a common-support condition for the instrument.

\begin{assumption}\label{assumption: common support in late}
There exists $\epsilon>0$ such that
\[
\epsilon<m^Z(X)<1-\epsilon.
\]
\end{assumption}

Define, for $z\in\{0,1\}$,
\begin{equation*}
\mu^Y_{Z=z}(X):=\operatorname{E}\{Y\mid Z=z,X\},
\qquad
\mu^D_{Z=z}(X):=\operatorname{E}\{D\mid Z=z,X\}.
\end{equation*}
Following the same logic as in Section \ref{sec: framework}, the null hypothesis can be represented through locally robust moments. Let
\begin{align}
U^{LATE}
:=&
V^Y-\delta_0 V^D,
\label{eq:U_Late}
\\
V^Y
:=&
\mu^Y_{Z=1}(X)-\mu^Y_{Z=0}(X)
+
[Y-\mu^Y_{Z=1}(X)]
\frac{Z}{m^Z(X)}
-
[Y-\mu^Y_{Z=0}(X)]
\frac{1-Z}{1-m^Z(X)},
\label{eq:VY_Late}
\\
V^D
:=&
\mu^D_{Z=1}(X)-\mu^D_{Z=0}(X)
+
[D-\mu^D_{Z=1}(X)]
\frac{Z}{m^Z(X)}
-
[D-\mu^D_{Z=0}(X)]
\frac{1-Z}{1-m^Z(X)}.
\label{eq:VD_Late}
\end{align}
Here, $V^Y$ is the locally robust analogue of the reduced-form effect of the instrument on the outcome, while $V^D$ is the corresponding first-stage effect on treatment take-up. The null hypothesis is equivalently characterized by
\begin{equation*}
\mathcal{H}_0^{LATE}
\Leftrightarrow
\operatorname{E}\{U^{LATE}\varphi(t^TX_c)\}
=
0
\qquad
\forall t\in\mathcal{T}.
\end{equation*}

We estimate $m^Z$, $\mu^Y_{Z=z}$, and $\mu^D_{Z=z}$ using the same cross-fitting procedure as in Section \ref{sec: test statistic}. Given the fitted values
\[
\left\{
\widehat \mu^Y_{Z=1,K(-i)}(X_i),
\widehat \mu^Y_{Z=0,K(-i)}(X_i),
\widehat \mu^D_{Z=1,K(-i)}(X_i),
\widehat \mu^D_{Z=0,K(-i)}(X_i),
\widehat m^Z_{K(-i)}(X_i)
\right\}_{i=1}^n,
\]
define
\begin{align}
\widehat U^{LATE}_i
:=&
\widehat V^Y_i
-
\widehat\delta^{LATE}\widehat V^D_i,
\label{eq:Uhat_Late}
\\
\widehat V^Y_i
:=&
\widehat\mu^Y_{Z=1,K(-i)}(X_i)
-
\widehat\mu^Y_{Z=0,K(-i)}(X_i)
\label{eq:VYhat_Late}
\\
&+
[Y_i-\widehat\mu^Y_{Z=1,K(-i)}(X_i)]
\frac{Z_i}{\widehat m^Z_{K(-i)}(X_i)}
-
[Y_i-\widehat\mu^Y_{Z=0,K(-i)}(X_i)]
\frac{1-Z_i}{1-\widehat m^Z_{K(-i)}(X_i)},
\nonumber
\\
\widehat V^D_i
:=&
\widehat\mu^D_{Z=1,K(-i)}(X_i)
-
\widehat\mu^D_{Z=0,K(-i)}(X_i)
\label{eq:VDhat_Late}
\\
&+
[D_i-\widehat\mu^D_{Z=1,K(-i)}(X_i)]
\frac{Z_i}{\widehat m^Z_{K(-i)}(X_i)}
-
[D_i-\widehat\mu^D_{Z=0,K(-i)}(X_i)]
\frac{1-Z_i}{1-\widehat m^Z_{K(-i)}(X_i)},
\nonumber
\\
\widehat\delta^{LATE}
:=&
\frac{\operatorname{E}_n\{\widehat V^Y_i\}}
{\operatorname{E}_n\{\widehat V^D_i\}}.
\label{eq:deltahat_Late}
\end{align}
The estimator $\widehat\delta^{LATE}$ is the locally robust analogue of the unconditional LATE estimator. It equals the ratio of the reduced-form and first-stage components, while allowing both components to be estimated flexibly.

The resulting test statistic is
\begin{equation}
\label{eq: test statistic late}
S_n^{LATE}
=
\int
\left|
\sqrt{n}
\operatorname{E}_n
\{\widehat U_i^{LATE}\phi_t(X_{c,i})\}
\right|^2
d\rho(t),
\end{equation}
where $\phi_t(X_c):=\varphi(t^TX_c)$. As before, choosing
\begin{equation*}
\phi_t(\cdot)=\exp(\bm{i}t^T\cdot)
\end{equation*}
yields the closed-form expression
\begin{equation*}
S_n^{LATE}
=
\frac{1}{n}
(\bm{\widehat U}^{LATE})^T
\bm{F_\rho}
\bm{\widehat U}^{LATE},
\end{equation*}
where
\begin{equation*}
\bm{\widehat U}^{LATE}
:=
(\widehat U^{LATE}_1,\ldots,\widehat U^{LATE}_n)^T.
\end{equation*}

We impose the following regularity conditions. Let $\zeta:=(Y,D,Z,X)$, let $P$ denote the distribution of $\zeta$, and let $\mathcal X$ denote the support of $X$. The norm $\|\cdot\|_{L^2(P)}$ is defined as in Section \ref{sec: test statistic}.

\begin{assumption}\label{as: iid , finite second moments, phi, mu late}
~\\
(i) $\{Y_i,D_i,Z_i,X_i\}_{i=1}^n$ is an iid sample;
\\
(ii) $Y$ is bounded;
\\
(iii) $\mathcal{T}$ is a compact subset of $\mathbb{R}^c$ containing a neighborhood of the origin and $X_c$ is bounded;
\\
(iv) $\varphi$ is an analytic non-polynomial function with $\partial^l\varphi(0)\neq 0$ for all $l\in\mathbb{N}$.
\end{assumption}

\begin{assumption}\label{as: convergence rates for ml estimators late}
~\\
(i) For any $j\in\{1,\ldots,N\}$,
\begin{align*}
&
\|
\widehat m^Z_{K-K_j}(X)-m^Z(X)
\|_{L^2(P)}
+
\|
\widehat \mu^Y_{Z=0,K-K_j}(X)
-
\mu^Y_{Z=0}(X)
\|_{L^2(P)}
\\
&+
\|
\widehat \mu^Y_{Z=1,K-K_j}(X)
-
\mu^Y_{Z=1}(X)
\|_{L^2(P)}
+
\|
\widehat \mu^D_{Z=0,K-K_j}(X)
-
\mu^D_{Z=0}(X)
\|_{L^2(P)}
\\
&+
\|
\widehat \mu^D_{Z=1,K-K_j}(X)
-
\mu^D_{Z=1}(X)
\|_{L^2(P)}
=
o_P(n^{-1/4});
\end{align*}

(ii) There exists a constant $C>0$ such that for any $j\in\{1,\ldots,N\}$,
\begin{align*}
\lim_n
\P\Bigg(
&
\sup_{x\in\mathcal X}
|\widehat\mu^D_{Z=0,K-K_j}(x)|
+
\sup_{x\in\mathcal X}
|\widehat\mu^D_{Z=1,K-K_j}(x)|
\\
&+
\sup_{x\in\mathcal X}
|\widehat\mu^Y_{Z=0,K-K_j}(x)|
+
\sup_{x\in\mathcal X}
|\widehat\mu^Y_{Z=1,K-K_j}(x)|
<C
\Bigg)
=
1;
\end{align*}

(iii) There exists $\epsilon>0$ such that for any $j\in\{1,\ldots,N\}$,
\begin{equation*}
\lim_n
\P\left(
\epsilon
<
\inf_{x\in\mathcal X}
\widehat m^Z_{K-K_j}(x)
\right)
=
\lim_n
\P\left(
\sup_{x\in\mathcal X}
\widehat m^Z_{K-K_j}(x)
<
1-\epsilon
\right)
=
1.
\end{equation*}
\end{assumption}

Assumptions \ref{as: iid , finite second moments, phi, mu late} and \ref{as: convergence rates for ml estimators late} are the LATE analogues of the regularity conditions used above. They ensure that the reduced-form and first-stage nuisance functions are estimated accurately enough and that the estimated instrument propensity score remains bounded away from zero and one.

\begin{tcolorbox}[colback=gray!6!white, colframe=black]
\begin{proposition}\label{prop: late statistics asymptotics}
Let Assumptions \ref{as: independence in late}, \ref{as: monotonicity in late}, \ref{assumption: conditional relevance late}, \ref{assumption: common support in late}, \ref{as: iid , finite second moments, phi, mu late}, and \ref{as: convergence rates for ml estimators late} hold. Then:

(a) For
\begin{equation*}
f_t(\zeta)
:=
U^{LATE}\phi_t(X_c)
-
[\operatorname{E}\{V^D\phi_t(X_c)\}]
\frac{
V^Y-\operatorname{E}\{V^Y\}
-
\delta_0\left(V^D-\operatorname{E}\{V^D\}\right)
}
{
\operatorname{E}\{\mu^D_{Z=1}(X)-\mu^D_{Z=0}(X)\}
},
\end{equation*}
under $\mathcal{H}_0^{LATE}$,
\begin{equation*}
S_n^{LATE}
\overset{d}{\to}
\int
|\mathbb{G}(f_t)|^2
d\rho(t),
\end{equation*}
where $\{\mathbb{G}(f_t):t\in\mathcal{T}\}$ is a zero-mean tight Gaussian process taking values in $\ell^\infty(\mathcal{T})$ and characterized by the collection of covariances
\begin{equation*}
\{
\operatorname{E}f_{t_1}(\zeta)f_{t_2}(\zeta)
:
t_1,t_2\in\mathcal{T}
\}.
\end{equation*}

(b) Under $\mathcal{H}_1^{LATE}$,
\begin{equation*}
S_n^{LATE}
\overset{P}{\to}
\infty.
\end{equation*}
\end{proposition}
\end{tcolorbox}

Part (a) establishes the asymptotic distribution of the test statistic under the null hypothesis of homogeneous local treatment effects. Part (b) shows that the statistic diverges under the alternative, implying consistency against fixed alternatives. The proof is provided in the supplementary material.

We obtain critical values using a bootstrap procedure analogous to that in Section \ref{sec: bootstrap test}. The nuisance functions are estimated only once and are kept fixed across bootstrap iterations. Since the LATE parameter is a ratio, the bootstrap recomputes the reduced-form and first-stage components in each iteration. Let $\{\xi_i\}_{i=1}^n$ be an iid sequence of bootstrap weights independent from the sample data, with $\operatorname{E}\{\xi\}=\operatorname{Var}\{\xi\}=1$. We define 
\begin{gather*}
    \widehat \delta^{LATE*}:=\frac{\operatorname{E}_n\{\xi_i \widehat V^Y_i\}}{\operatorname{E}_n\{\xi_i \widehat V^D_i\}}\,,\quad
    \widehat U^{LATE*}_i:=\xi_i[\widehat V_i^Y - \widehat \delta^{LATE*} \widehat V_i^D]\, .
\end{gather*}
The bootstrap counterpart of the test statistic is 
\begin{equation*}
    S_n^{LATE*}:=\int_{ }\left|\sqrt{n} \operatorname{E}_n\{ [\widehat U_i^{LATE*}-\widehat U_i^{LATE}]\phi_t(X_{c,i})\}\right|^2 \mu(dt)\, .
\end{equation*}
By setting $\phi_t(\cdot)=\exp(\textbf{i}\cdot)$, the above statistic has the following simple expression
\begin{equation*}
    S_n^{LATE*}=\frac{1}{n}(\bm{\widehat U^{LATE*}}-\bm{\widehat U^{LATE}})^T \bm{F_\rho} (\bm{\widehat U^{LATE*}}-\bm{\widehat U^{LATE}})\,,
\end{equation*}
where $\bm{\widehat U^{LATE*}}:=(\widehat U_1^{LATE*},\ldots,\widehat U_n^{LATE*})^T$ and $\bm{\widehat U^{LATE}}:=(\widehat U_1^{LATE},\ldots,\widehat U_n^{LATE})^T$. In practice, to implement the bootstrap test, we can use the Monte Carlo procedure described in Section \ref{sec: bootstrap test}.

Let $\widehat q_{1-\alpha}^{LATE}$ denote the empirical $(1-\alpha)$ quantile of the bootstrap statistics generated by the procedure above. The following proposition establishes the validity of this LATE bootstrap critical value.

\begin{tcolorbox}[colback=gray!6!white, colframe=black]
\begin{proposition}\label{prop: bootstrap validity late}
Let Assumptions \ref{as: independence in late}, \ref{as: monotonicity in late}, \ref{assumption: conditional relevance late}, \ref{assumption: common support in late}, \ref{as: iid , finite second moments, phi, mu late}, and \ref{as: convergence rates for ml estimators late} hold. Then:

(a) Under $\mathcal{H}_0^{LATE}$,
\begin{equation*}
\Pr(S_n^{LATE}>\widehat q_{1-\alpha}^{LATE})
\rightarrow
\alpha.
\end{equation*}

(b) Under $\mathcal{H}_1^{LATE}$,
\begin{equation*}
\Pr(S_n^{LATE}>\widehat q_{1-\alpha}^{LATE})
\rightarrow
1.
\end{equation*}
\end{proposition}
\end{tcolorbox}

Proposition \ref{prop: bootstrap validity late} establishes that the bootstrap test has asymptotically correct size under the null hypothesis and is consistent under fixed alternatives. The proof is provided in the supplementary material.

\section{Simulations}
\label{sec: simulations did}

This section evaluates the finite-sample performance of our bootstrap test for DiD with repeated cross-sections. Appendix \ref{sec: ate simulations} provides additional simulations for the ATE under conditional unconfoundedness.

\subsection{Data-generating Process}

We use a data-generating process (DGP) similar to that in \citet{chang2020double}, adapted to feature conditional, but not unconditional, parallel trends, different levels of treatment-effect heterogeneity, and common support.

We first simulate $X\sim \mathcal{N}(0,I_{10})$ and set $\theta_1=\left(1,\frac{1}{2},\frac{1}{3},\frac{1}{4},\frac{1}{5},0,0,0,0,0\right)'$. We then define
\begin{align*}
    &m^G(X_i)=\left(1+\exp\left(\frac{-X_i'\theta_1}{2\sqrt{5}}\right)\right)^{-1},\\
    &G_i|X_i\sim \operatorname{Bern}\left(m^G(X_i)\right),\\
    &T_i|X_i\sim \operatorname{Bern}\left(1/2\right).
\end{align*}
Thus, assignment to the treated group depends on the first five covariates, with the first covariate receiving the largest weight. By contrast, whether a unit is observed in the post-treatment period is independent of its covariates.

Next, we generate $\varepsilon_{i,k} \sim \mathcal{N}(0,0.01)$ for $k \in \{1,2,3\}$ and set $\theta_2=\left(\frac{3}{2},1,\frac{5}{6},\frac{3}{4},\frac{7}{10},0,0,0,0,0\right)'$. Potential and observed outcomes are generated as follows:
\begin{align*}
    &Y_{i,0}(0)=X_i'\theta_2+\varepsilon_{i,1},\\
    &Y_{i,1}(0)=1+(1+\xi)Y_{i,0}(0)+\varepsilon_{i,2},\\
    &Y_{i,1}(1)=3+(1+\gamma)Y_{i,1}(0)+\varepsilon_{i,3},\\
    &Y_i=Y_{i,0}(0)(1-T_i)+\big(Y_{i,1}(0)(1-G_i)+Y_{i,1}(1)G_i\big)T_i.
\end{align*}

When $\xi\neq 0$, the parallel-trends assumption holds only conditionally. We therefore fix $\xi=1/2$. The parameter $\gamma$ controls the degree of heterogeneity in the ATT. In particular, for any subset $X_c$, if $\gamma=0$, then
\[
\E\left[Y_{i,1}(1)- Y_{i,1}(0)\mid G_i=1, X_{c,i}\right]=3,
\]
so the null hypothesis of no ATT heterogeneity with respect to $X_c$ holds. If $\gamma\neq 0$ and $\theta_{2,c}$ denotes the subvector of $\theta_2$ corresponding to the covariates in $X_c$, then
\[
\E\left[Y_{i,1}(1)- Y_{i,1}(0)\mid G_i=1, X_{c,i}\right]
=
3 + \gamma\left(1+1.5X_{c,i}'\theta_{2,c}\right).
\]
Thus, whenever $X_c$ contains at least one of the first five covariates in $X$, the null hypothesis is violated.

\subsection{Performance}

We evaluate rejection rates under the null and alternative hypotheses across 1,000 Monte Carlo simulations. We consider sample sizes $n\in\{500,1000\}$ and use $B=499$ bootstrap iterations. We test for heterogeneity with respect to subsets $X_c$ containing the first $d\in\{2,5\}$ components of $X$, that is, $X_c=(X_1,X_2)$ or $X_c=(X_1,\ldots,X_5)$. For nuisance estimation, we use lasso and random forests. Cross-fitting is implemented with five equally sized random folds. We set $\phi_t(\cdot)=\exp(\bm{i}t^\top\cdot)$ and let $\rho(\cdot)$ be the standard Gaussian density, so that $\mathcal{F}_\rho$ corresponds to a Gaussian kernel with bandwidth $h=1$.

\begin{table}[htb]
	\caption{Rejection Rate under the Null Hypothesis}
	\label{tb:null_hyp_DiD}
    \centering
    \begin{adjustbox}{max width=\textwidth, max totalheight=\textheight}
	\begin{threeparttable}
\begin{tabular}{llSSSS}   
 \toprule\toprule
  &&   \multicolumn{2}{c}{$\alpha=5\%$}&\multicolumn{2}{c}{$\alpha=10\%$} \\\cmidrule(lr){3-4}\cmidrule(lr){5-6}
$n$&Method & \multicolumn{1}{c}{$d=2$} & \multicolumn{1}{c}{$d=5$} & \multicolumn{1}{c}{$d=2$}& \multicolumn{1}{c}{$d=5$}\\ 
  \midrule
 500 & Lasso & 0.052 & 0.036 & 0.095 & 0.079\\
     & RF   & 0.051 & 0.046 & 0.103 & 0.091\\
  \midrule
1000 & Lasso & 0.052 & 0.043 & 0.105 & 0.092\\
     & RF    & 0.054 & 0.052 & 0.104 & 0.098\\
   \bottomrule\bottomrule
\end{tabular}
\begin{tablenotes}[flushleft,para]
\footnotesize
Note: This table reports rejection rates across 1,000 Monte Carlo simulations under the null hypothesis for the bootstrap test in the DiD model with repeated cross-sections. The sample size is $n\in\{500, 1000\}$. Nuisance functions are estimated using either lasso or random forests. Cross-fitting is performed with five folds. For each Monte Carlo iteration, we use 499 bootstrap replications. We report rejection rates for significance levels $\alpha\in\{5\%,10\%\}$ and for different dimensions $d \in \{2,5 \}$ of the vector $X_c$. The Fourier matrix is computed using a Gaussian kernel with bandwidth $1$.
\end{tablenotes}
\end{threeparttable}
\end{adjustbox}
\end{table}

We first evaluate the validity of the bootstrap test under the null hypothesis, corresponding to $\gamma=0$. Table \ref{tb:null_hyp_DiD} reports rejection rates under the null for significance levels of 5\% and 10\%. For $d=2$, the test is close to correctly sized. For $d=5$, it becomes slightly conservative. This pattern is stable across sample sizes and nuisance-estimation methods.

\begin{figure}[htb]
    \centering
    \caption{Rejection Rates across Heterogeneity Levels}
    \label{fig:Chang}
    \includegraphics[width=\linewidth]{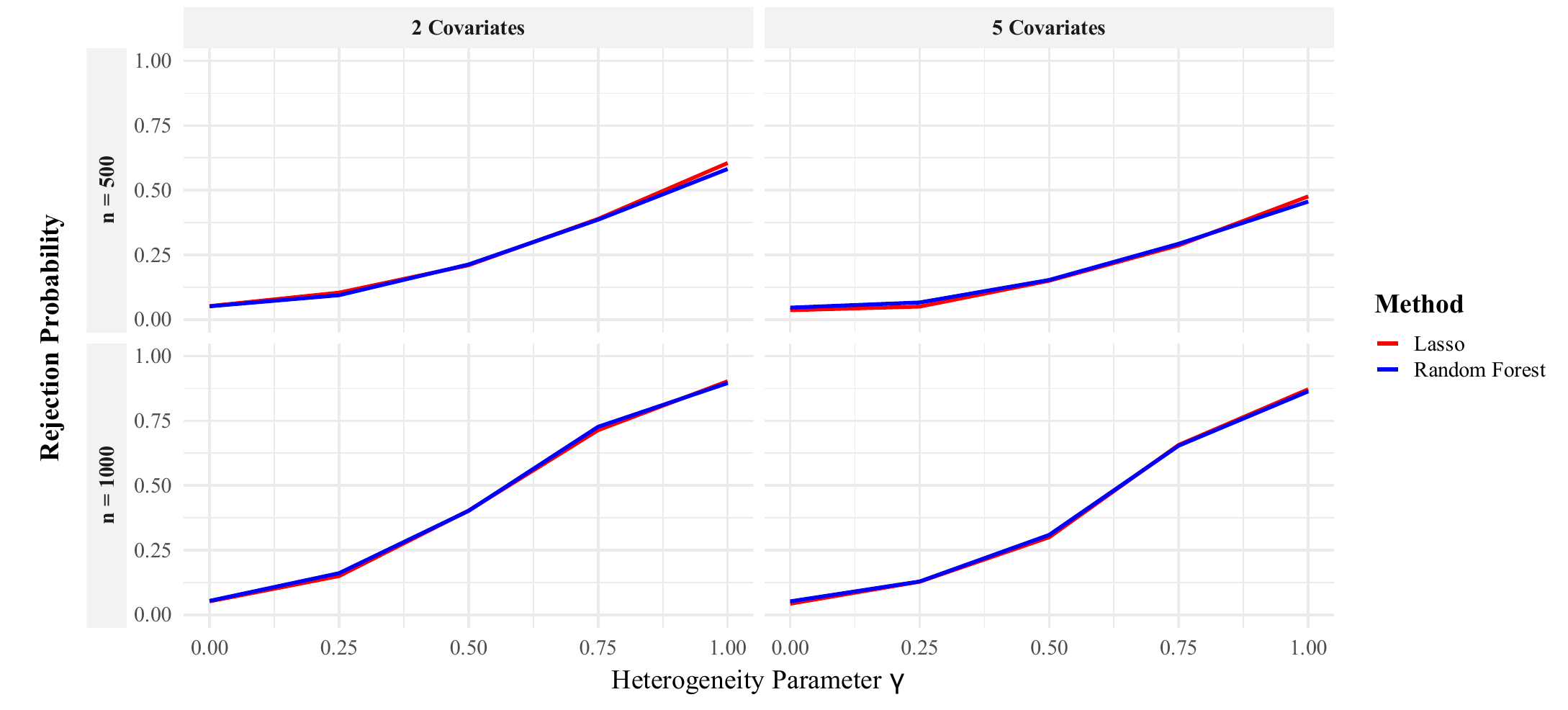}
    \floatfoot{\justifying Note: This figure reports rejection rates at the 5\% significance level for different values of $\gamma$, which controls heterogeneity in the ATT, for the bootstrap test in the DiD model with repeated cross-sections. Rejection rates are computed from 1,000 Monte Carlo simulations. The sample size is $n\in\{500,1000\}$. Nuisance functions are estimated using either lasso or random forests. Cross-fitting is performed with five folds, and 499 bootstrap replications are used in each iteration. We consider different dimensions $d \in \{2,5 \}$ of the vector $X_c$. The Fourier matrix is computed using a Gaussian kernel with bandwidth $1$.}
\end{figure}

We next examine power under the alternative by evaluating rejection rates for different values of $\gamma$. Figure \ref{fig:Chang} displays the corresponding power curves at the 5\% significance level. As expected, rejection rates increase with $\gamma$ and approach one. Power is lower for $n=500$, but including more heterogeneity covariates, $d=5$, does not substantially affect rejection rates. Finally, lasso and random forests yield nearly identical results.

\section{Empirical Applications \label{sec: empirical applications}}

We illustrate the proposed test using two established empirical applications. The first studies heterogeneity in the effect of 401(k) eligibility on household financial assets under conditional unconfoundedness. The second studies heterogeneity in the effect of tariff reductions on corruption in a difference-in-differences setting with repeated cross-sections. In both applications, our goal is not to revisit the original identification arguments in detail, but to demonstrate how the proposed test can be implemented with flexible ML-based nuisance estimation.

\subsection{401(k) Eligibility and Financial Assets}

The first application considers the effect of eligibility for a 401(k) retirement savings plan on household financial asset accumulation. We follow the empirical setting of \citet{poterba_401k_1995}, which was later revisited in the double ML framework of \citet{chernozhukov2018double}. Eligibility for a 401(k) plan is not randomly assigned, since workers employed at firms offering such plans may differ systematically from other workers. Following this literature, we treat eligibility as conditionally exogenous after controlling for income and demographic characteristics related to job choice and saving behavior.

The data come from the 1991 Survey of Income and Program Participation (SIPP), using the sample analyzed in \citet{chernozhukov2018double}. The outcome is net financial assets, defined as household financial assets net of non-mortgage debt, and the treatment is an indicator for 401(k) eligibility. The covariates include age, income, education, family size, marital status, two-earner status, defined benefit pension status, IRA participation, and home ownership. We estimate the nuisance functions using lasso and random forests.

\begin{figure}[htb]
    \centering
    \caption{Heterogeneity Test: 401(k) Eligibility and Financial Assets}
    \label{fig:app1}
    \includegraphics[width=0.7\linewidth]{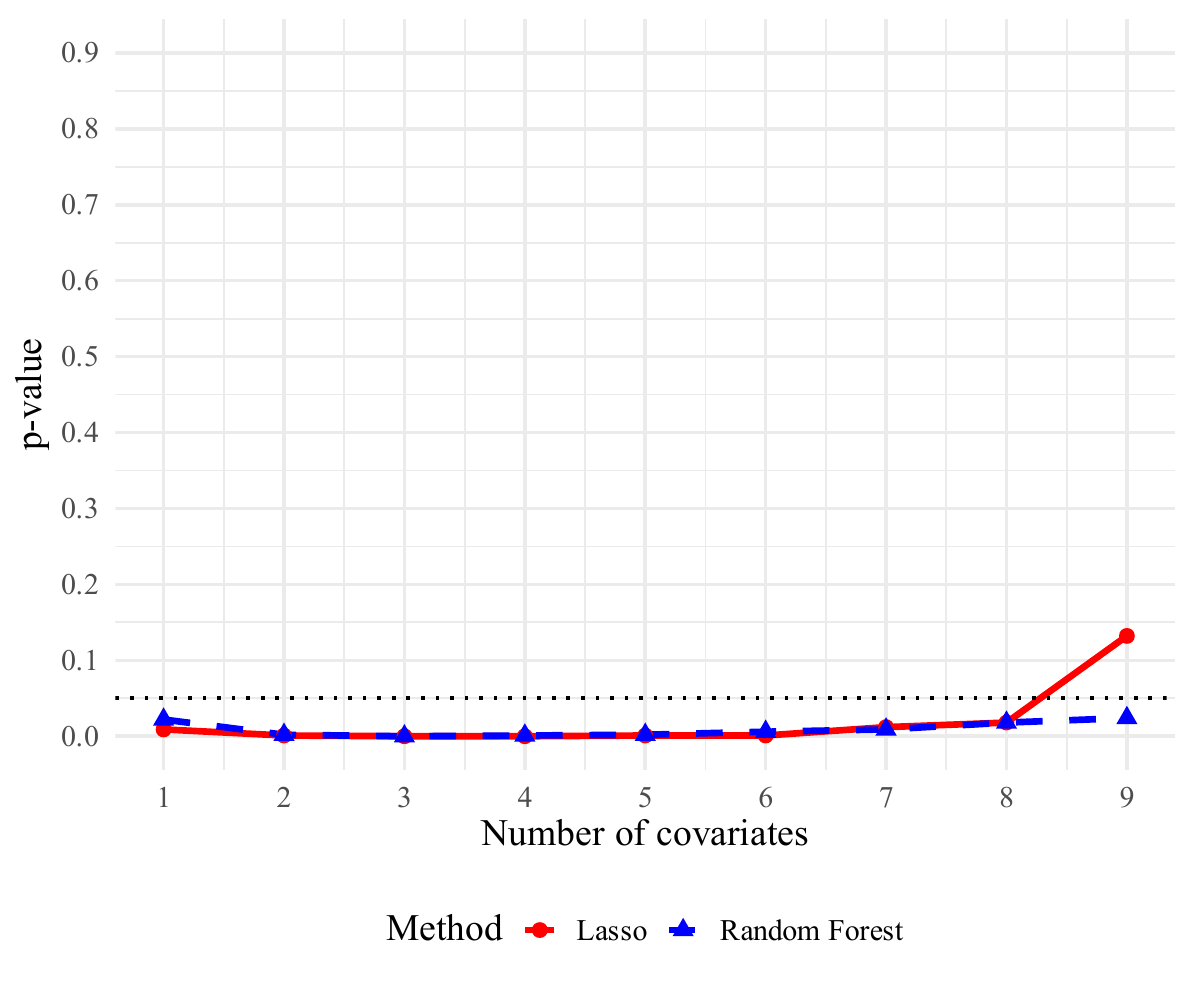}
    \floatfoot{\justifying Note: The figure reports p-values from the proposed heterogeneity test for the effect of 401(k) eligibility on net financial assets. The heterogeneity variables are added sequentially using observable household characteristics. Nuisance functions are estimated using lasso and random forests. The Fourier matrix is computed using a Gaussian kernel.}
\end{figure}

Figure~\ref{fig:app1} reports p-values as the set of heterogeneity variables is expanded sequentially, beginning with age and income and then adding further demographic and financial characteristics. The results provide evidence of treatment-effect heterogeneity across observable household characteristics for both nuisance-estimation methods. The p-values are generally smaller in lower-dimensional specifications and increase as more heterogeneity variables are included, especially with lasso. This pattern is consistent with the simulation evidence, where increasing the dimension of $X_c$ reduces power. The random forest results are more stable across specifications.

The findings are economically plausible. The effect of access to a tax-advantaged retirement account may differ by income, age, and pre-existing savings opportunities. The application therefore illustrates how the proposed procedure can detect systematic heterogeneity in an observational setting while allowing for flexible adjustment for confounding.

\subsection{Trade Liberalization and Corruption}

The second application studies treatment-effect heterogeneity in the difference-in-differences setting analyzed by \citet{sequeira_corruption_2016}. The application examines how reductions in import tariffs affected corruption along the South Africa--Mozambique trade corridor. The empirical design exploits the Southern African Development Community (SADC) Trade Protocol, which gradually reduced tariffs on South African imports into Mozambique and generated a large tariff reduction in 2008.

We use the sample from \citet{chang2020double}, which combines disaggregated customs data with audit data on shipments passing through the port of Maputo and the Mozambique--South Africa border crossing. The outcome is the log bribe amount associated with a shipment. Treatment is defined by whether a product category experienced a tariff reduction in 2008, and the post-treatment indicator captures the period after the reform. The covariates include shipment characteristics, product categories, inspection indicators, shipment values, tariff levels, and other controls related to the import process. We estimate the nuisance functions using lasso and random forests in the repeated-cross-section DiD framework.

\begin{figure}[htb]
    \centering
    \caption{Heterogeneity Test: Trade Liberalization and Corruption}
    \label{fig:app2}
    \includegraphics[width=0.7\linewidth]{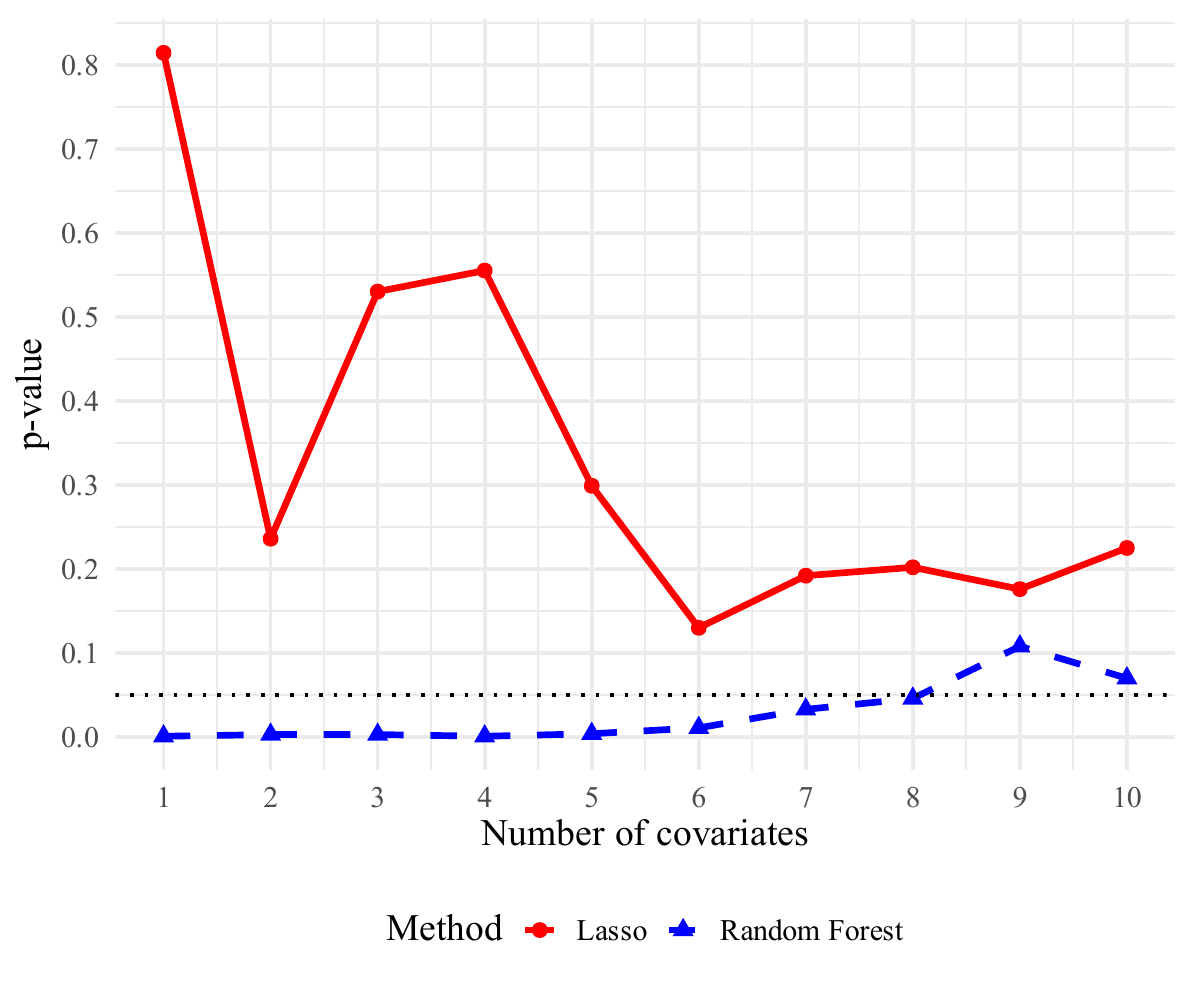}
    \floatfoot{\justifying Note: The figure reports p-values from the proposed heterogeneity test for the effect of tariff reductions on log bribe payments. The heterogeneity variables are added sequentially using observable shipment and product characteristics. Nuisance functions are estimated using lasso and random forests. The Fourier matrix is computed using a Gaussian kernel.}
\end{figure}

Figure~\ref{fig:app2} reports p-values as the set of heterogeneity variables is expanded sequentially. The random forest specification detects significant heterogeneity across several specifications, whereas the lasso specification generally does not. This difference suggests that flexible nuisance estimation may be important in this application, where the relevant adjustment functions may involve nonlinearities or interactions. As in the 401(k) application, p-values tend to increase as the dimension of $X_c$ grows, in line with the finite-sample patterns documented in the simulations.

The evidence is consistent with economically meaningful heterogeneity. The effect of tariff reductions on bribe payments may vary with shipment characteristics, product types, tariff exposure, and monitoring intensity. The application shows that the proposed test can be implemented in repeated-cross-section DiD designs and can reveal systematic heterogeneity while preserving the use of flexible ML-based nuisance estimation.

\section{Conclusions}

This paper develops a nonparametric omnibus test for assessing whether treatment effects vary systematically with observed characteristics. The procedure is designed for settings in which identification may rely on a rich, potentially high-dimensional set of controls, while heterogeneity is evaluated with respect to a lower-dimensional subset of covariates. The test is compatible with flexible nuisance estimation, including modern ML methods, and is based on locally robust moments that reduce sensitivity to first-stage estimation error.

The framework applies to several common empirical designs. We first develop the test for average treatment effects under conditional unconfoundedness and then extend it to treatment effects on the treated in difference-in-differences designs and to local average treatment effects in instrumental-variables settings. For each design, inference is implemented using a computationally efficient bootstrap that keeps the nuisance estimates fixed across bootstrap iterations.

Monte Carlo simulations show that the proposed test attains near-nominal size under the null and has good power against heterogeneous alternatives. The empirical applications illustrate how the procedure can be used to detect systematic treatment-effect heterogeneity in observational and difference-in-differences settings with flexible ML-based nuisance estimation.

Several extensions are left for future work. In particular, it would be useful to study settings in which both identification and heterogeneity are governed by high-dimensional covariates, as well as data structures beyond the designs considered here. Overall, the proposed method provides a practical tool for researchers who want to formally assess whether treatment effects differ across observable and policy-relevant characteristics.

\clearpage
\bibliographystyle{apalike}
\bibliography{biblio}

\begin{thebibliography}{}

\bibitem[Abadie, 2005]{abadie2005semiparametric}
Abadie, A. (2005).
\newblock Semiparametric difference-in-differences estimators.
\newblock {\em The review of economic studies}, 72(1):1--19.

\bibitem[Athey and Imbens, 2016]{ath2016}
Athey, S. and Imbens, G. (2016).
\newblock Recursive partitioning for heterogeneous causal effects.
\newblock {\em Proceedings of the National Academy of Sciences},
  113(27):7353--7360.

\bibitem[Athey and Imbens, 2019]{athey2019machine}
Athey, S. and Imbens, G.~W. (2019).
\newblock Machine learning methods that economists should know about.
\newblock {\em Annual Review of Economics}, 11(1):685--725.

\bibitem[Athey et~al., 2019]{ath2019}
Athey, S., Tibshirani, J., and Wager, S. (2019).
\newblock Generalized random forests.
\newblock {\em Annals of Statistics}, 47(2):1148--1178.

\bibitem[Bierens, 2016]{bierens2016econometric}
Bierens, H.~J. (2016).
\newblock {\em Econometric Model Specification}.
\newblock World Scientific.

\bibitem[Chang, 2020]{chang2020double}
Chang, N.-C. (2020).
\newblock Double/debiased machine learning for difference-in-differences
  models.
\newblock {\em The Econometrics Journal}, 23(2):177--191.

\bibitem[Chernozhukov et~al., 2018]{chernozhukov2018double}
Chernozhukov, V., Chetverikov, D., Demirer, M., Duflo, E., Hansen, C., Newey,
  W., and Robins, J. (2018).
\newblock Double/debiased machine learning for treatment and structural
  parameters.

\bibitem[Chernozhukov et~al.,
  2025]{chernozhukov2023fisherschultzlecturegenericmachine}
Chernozhukov, V., Demirer, M., Duflo, E., and Fernández-Val, I. (2025).
\newblock {Fisher-Schultz Lecture: Generic} machine learning inference on
  heterogenous treatment effects in randomized experiments, with an application
  to immunization in {India}.
\newblock {\em Econometrica}, 93(4):1121--1164.

\bibitem[Chernozhukov et~al., 2024]{chern2024}
Chernozhukov, V., Hansen, C., Kallus, N., Spindler, M., and Syrgkanis, V.
  (2024).
\newblock Applied causal inference powered by {ML} and {AI}.
\newblock {\em arXiv:2403.02467}.

\bibitem[Chernozhukov et~al., 2022]{chernozhukov2018}
Chernozhukov, V., Newey, W.~K., and Singh, R. (2022).
\newblock De-biased machine learning of global and local parameters using
  regularized riesz representers.
\newblock {\em The Econometrics Journal}, 25(3):576--601.

\bibitem[Chernozhukov et~al., 2023]{chernozhukov2021}
Chernozhukov, V., Newey, W.~K., and Singh, R. (2023).
\newblock A simple and general debiased machine learning theorem with
  finite-sample guarantees.
\newblock {\em Biometrika}, 110(1):257--264.

\bibitem[Crump et~al., 2008]{crump2008nonparametric}
Crump, R.~K., Hotz, V.~J., Imbens, G.~W., and Mitnik, O.~A. (2008).
\newblock Nonparametric tests for treatment effect heterogeneity.
\newblock {\em The Review of Economics and Statistics}, 90(3):389--405.

\bibitem[Ding et~al., 2016]{ding2016randomization}
Ding, P., Feller, A., and Miratrix, L. (2016).
\newblock Randomization inference for treatment effect variation.
\newblock {\em Journal of the Royal Statistical Society Series B: Statistical
  Methodology}, 78(3):655--671.

\bibitem[Ding et~al., 2019]{ding2019decomposing}
Ding, P., Feller, A., and Miratrix, L. (2019).
\newblock Decomposing treatment effect variation.
\newblock {\em Journal of the American Statistical Association},
  114(525):304--317.

\bibitem[Escanciano, 2026]{escanciano2026kernel}
Escanciano, J.~C. (2026).
\newblock Kernel-based specification testing with high-dimensional nuisance
  parameters.
\newblock Working paper, presented at the ISNPS Conference, June 2026.

\bibitem[Fan et~al., 2022]{fan2022}
Fan, Q., Hsu, Y.-C., Lieli, R.~P., and Zhang, Y. (2022).
\newblock Estimation of conditional average treatment effects with
  high-dimensional data.
\newblock {\em Journal of Business \& Economic Statistics}, 40(1):313--327.

\bibitem[Farrell et~al., 2021]{farrell2021deep}
Farrell, M.~H., Liang, T., and Misra, S. (2021).
\newblock Deep neural networks for estimation and inference.
\newblock {\em Econometrica}, 89(1):181--213.

\bibitem[Foster and Syrgkanis, 2023]{Foster2019}
Foster, D.~J. and Syrgkanis, V. (2023).
\newblock Orthogonal statistical learning.
\newblock {\em Annals of Statistics}, 51(3):879--908.

\bibitem[Friedberg et~al., 2021]{Fried21}
Friedberg, R., Tibshirani, J., Athey, S., and Wager, S. (2021).
\newblock Local linear forests.
\newblock {\em Journal of Computational and Graphical Statistics},
  30(2):503--517.

\bibitem[Green and Kern, 2012]{green2012}
Green, D.~P. and Kern, H.~L. (2012).
\newblock Modeling heterogeneous treatment effects in survey experiments with
  bayesian additive regression trees.
\newblock {\em Public Opinion Quarterly}, 76(3):491--511.

\bibitem[Hahn et~al., 2020]{hahn2020}
Hahn, P.~R., Murray, J.~S., and Carvalho, C. (2020).
\newblock Bayesian regression tree models for causal inference: Regularization,
  confounding, and heterogeneous effects.
\newblock {\em Bayesian Analysis}, 15(3):965--1056.

\bibitem[Hansen et~al., 2023]{Hansen2017}
Hansen, C., Kozbur, D., and Misra, S. (2023).
\newblock Targeted undersmoothing: Sensitivity analysis for sparse estimators.
\newblock {\em Review of Economics and Statistics}, 105(1):101--118.

\bibitem[Heckman et~al., 1997]{heck97}
Heckman, J., Smith, J., and Clements, N. (1997).
\newblock {Making the Most Out of Programme Evaluations and Social Experiments:
  Accounting for Heterogeneity in Programme Impacts}.
\newblock {\em Review of Economic Studies}, 64(4):487--535.

\bibitem[Imai and Li, 2025a]{Imai2025b}
Imai, K. and Li, M.~L. (2025a).
\newblock A comment on: “{Fisher–Schultz Lecture: Generic} machine learning
  inference on heterogeneous treatment effects in randomized experiments, with
  an application to immunization in {India” by Victor Chernozhukov, Mert
  Demirer, Esther Duflo, and Iván Fernández-Val}.
\newblock {\em Econometrica}, 93(4):1165--1170.

\bibitem[Imai and Li, 2025b]{Imai2025a}
Imai, K. and Li, M.~L. (2025b).
\newblock Statistical inference for heterogeneous treatment effects discovered
  by generic machine learning in randomized experiments.
\newblock {\em Journal of Business \& Economic Statistics}, 43(1):256--268.

\bibitem[Kennedy, 2023]{kennedy2023towards}
Kennedy, E.~H. (2023).
\newblock Towards optimal doubly robust estimation of heterogeneous causal
  effects.
\newblock {\em Electronic Journal of Statistics}, 17(2):3008--3049.

\bibitem[Lechner and Mareckova, 2022]{lech2022}
Lechner, M. and Mareckova, J. (2022).
\newblock Modified causal forest.
\newblock {\em arXiv:2209.03744}.

\bibitem[Lechner and Zimmert, 2019]{lech2019}
Lechner, M. and Zimmert, M. (2019).
\newblock Nonparametric estimation of causal heterogeneity under
  high-dimensional confounding.
\newblock {\em arXiv: 1908.08779}.

\bibitem[List et~al., 2025]{list2025}
List, J.~A., Muir, I., and Sun, G. (2025).
\newblock Using machine learning for efficient flexible regression adjustment
  in economic experiments.
\newblock {\em Econometric Reviews}, 44(1):2--40.

\bibitem[Nekipelov et~al., 2022]{nekipelov2022}
Nekipelov, D., Semenova, V., and Syrgkanis, V. (2022).
\newblock Regularised orthogonal machine learning for nonlinear semiparametric
  models.
\newblock {\em The Econometrics Journal}, 25(1):233--255.

\bibitem[Nie and Wager, 2020]{Nie2020}
Nie, X. and Wager, S. (2020).
\newblock Quasi-oracle estimation of heterogeneous treatment effects.
\newblock {\em Biometrika}, 108(2):299--–319.

\bibitem[Poterba et~al., 1995]{poterba_401k_1995}
Poterba, J.~M., Venti, S.~F., and Wise, D.~A. (1995).
\newblock Do 401(k) contributions crowd out other personal saving?
\newblock {\em Journal of Public Economics}, 58(1):1--32.

\bibitem[Scheidegger et~al., 2026]{scheidegger2026inference}
Scheidegger, C., Guo, Z., and B{\"u}hlmann, P. (2026).
\newblock Inference for heterogeneous treatment effects with efficient
  instruments and machine learning.
\newblock {\em Electronic Journal of Statistics}, 20(1):718--770.

\bibitem[Scornet et~al., 2015]{scornet2015consistency}
Scornet, E., Biau, G., and Vert, J.-P. (2015).
\newblock Consistency of random forests.
\newblock {\em Annals of Statistics}, 43(4).

\bibitem[Semenova and Chernozhukov, 2021]{semenova2021debiased}
Semenova, V. and Chernozhukov, V. (2021).
\newblock Debiased machine learning of conditional average treatment effects
  and other causal functions.
\newblock {\em The Econometrics Journal}, 24(2):264--289.

\bibitem[Semenova et~al., 2023]{semenova2017}
Semenova, V., Goldman, M., Chernozhukov, V., and Taddy, M. (2023).
\newblock Inference on heterogeneous treatment effects in high-dimensional
  dynamic panels under weak dependence.
\newblock {\em Quantitative Economics}, 14(2):471--510.

\bibitem[Sequeira, 2016]{sequeira_corruption_2016}
Sequeira, S. (2016).
\newblock Corruption, {Trade} {Costs}, and {Gains} from {Tariff}
  {Liberalization}: {Evidence} from {Southern} {Africa}.
\newblock {\em American Economic Review}, 106(10):3029--3063.

\bibitem[Tabord-Meehan, 2023]{tabord2023stratification}
Tabord-Meehan, M. (2023).
\newblock Stratification trees for adaptive randomisation in randomised
  controlled trials.
\newblock {\em Review of Economic Studies}, 90(5):2646--2673.

\bibitem[Taddy et~al., 2016]{tad2016}
Taddy, M., Gardner, M., Chen, L., and Draper, D. (2016).
\newblock A nonparametric bayesian analysis of heterogeneous treatment effects
  in digital experimentation.
\newblock {\em Journal of Business \& Economic Statistics}, 34(4):661--672.

\bibitem[Wager and Athey, 2018]{wager2018estimation}
Wager, S. and Athey, S. (2018).
\newblock Estimation and inference of heterogeneous treatment effects using
  random forests.
\newblock {\em Journal of the American Statistical Association},
  113(523):1228--1242.

\end{thebibliography}

\clearpage
\appendix
\renewcommand\thefigure{\thesection.\arabic{figure}} 
\renewcommand\thetable{\thesection.\arabic{table}} 
\setcounter{figure}{0}   
\setcounter{table}{0}

\section{Derivations of Equations \eqref{eq: equivalent null hypothesis in DiD} and \eqref{eq: equivalent null hypothesis in DiD rcs}}\label{sec: proof of H0DiD and H0DiDrcs}
To obtain the equivalence in \eqref{eq: equivalent null hypothesis in DiD}, define $\mu_1^{Y_1-Y_0}(X)=\operatorname{E}\{\Delta Y |X, G=1\}$ and recall that $\mu_0^{Y_1-Y_0}(X)=\operatorname{E}\{\Delta Y |X, G=0\}$. We have the following:
\begin{align*}\label{eq: reformulation of H0 did}
	\mathcal{H}_0^{DiD}&\overset{(a)}{\Leftrightarrow} \operatorname{E}\{[Y_{1}(1)-Y_{1}(0)] G |X_c\}=\delta_0 \operatorname{E}\{G|X_c\}\nonumber\\
	&\overset{(b)}{\Leftrightarrow} \operatorname{E}\Big\{\, \operatorname{E}\{[Y_{1}(1)-Y_{1}(0)] G\,|\,X\} \,\Big|\,X_c\Big\}=\delta_0 \operatorname{E}\{G\,|\,X_c\}\nonumber\\
	&\Leftrightarrow \operatorname{E}\Big\{\, \operatorname{E}\{Y_{1}(1)-Y_{1}(0)\,|G=1, X\} \,\operatorname{E}\{G|X\}\,\Big|\,X_c\Big\}=\delta_0 \operatorname{E}\{G\,|\,X_c\}\nonumber\\
	&\overset{(c)}{\Leftrightarrow} \operatorname{E}\Big\{ \operatorname{E}\{Y_{1}(1)-Y_{1}(0)\,|G=1, X\} \,m^G(X) - \delta_0 G \,\Big|\,X_c \Big\}=0\nonumber\\
	&\overset{(d)}{\Leftrightarrow}  \operatorname{E}\Big\{ [\mu^{Y_1-Y_0}_{1}(X) - \mu^{Y_1-Y_0}_{0}(X)] m^G(X) - \delta_0 G \Big|\,X_c\Big\}=0\nonumber\\
	&\overset{(e)}{\Leftrightarrow}  \operatorname{E}\Big\{ \operatorname{E}\{\Delta Y\, G|X\} -\operatorname{E}\{\Delta Y (1-G)|X\} \frac{m^G(X)}{1-m^G(X)} - \delta_0 G\,\Big|\,X_c\Big\}=0 \nonumber\\
	&\overset{(f)}{\Leftrightarrow} \operatorname{E}\Big\{\Delta Y\, G - \Delta Y(1-G) \frac{m^G(X)}{1-m^G(X)} - \delta_0 G \,\Big|\,X_c\Big\}=0\nonumber \\
	&\overset{(g)}{\Leftrightarrow} \operatorname{E}\Big\{\Delta Y \frac{[G-m^G(X)]}{1-m^G(X)} - \delta_0 G \,\Big|\,X_c\Big\}=0. 
\end{align*}
Equivalence $(a)$ is a direct consequence of $\mathcal{H}_0^{DiD}:\operatorname{E}\{Y_1(1)-Y_1(0)|G=1,X_c\}=\delta_0$ for some $\delta_0\in \mathbb{R}$.
Equivalence $(b)$ follows from the Law of Iterated Expectations, while equivalence $(c)$ uses the definition $m^G(X)=\operatorname{E}\{G|X\}$. For equivalence $(d)$, first observe that $\mu^{Y_1-Y_0}_{1}(X) - \mu^{Y_1-Y_0}_{0}(X)$ $=\operatorname{E}\{Y_1(1) - Y_0(0)|G=1,X\}$ $-\operatorname{E}\{Y_1(0) - Y_0(0)|G=0,X\}$ $=\operatorname{E}\{Y_1(1) - Y_1(0)|G=1,X\}$ $+ \operatorname{E}\{Y_1(0) - Y_0(0)|G=1,X\}$ $-\operatorname{E}\{Y_1(0) - Y_0(0)|G=0,X\}$ $=\operatorname{E}\{Y_1(1) - Y_1(0)|G=1,X\}$, where the last equality follows from Assumption \ref{as: coditional common trend in did}. Hence, equivalence $(d)$ follows. Equivalence $(e)$ follows from the definitions of $\mu^{Y_1-Y_0}_{0}(X)$ and $\mu^{Y_1-Y_0}_{1}(X)$. Equivalence $(f)$ follow from  the Law of Iterated Expectations. Equivalence $(g)$ follows from re-arranging terms. 
Finally, since 
\begin{equation*}
	\operatorname{E}\{U^{DiD}|X\}=\operatorname{E}\left\{\Delta Y \frac{G-m^G(X)}{1-m^G(X)}-\delta_0 G\Big |X\right\}\,
\end{equation*}
and $\operatorname{E}\{U^{DiD}|X_c\}=0\Leftrightarrow\operatorname{E}\{U^{DiD} \varphi(t^T X_c)\}=0\,\forall t \in\mathcal{T}$,
Equation \eqref{eq: equivalent null hypothesis in DiD} follows.\\

We now derive Equation \eqref{eq: equivalent null hypothesis in DiD rcs}.
Under Assumptions \ref{as: coditional common trend in did} and \ref{as: support of the treated in did}, \cite{abadie2005semiparametric} identifies the conditional treatment effect on the treated units, $\operatorname{E}\{Y_{i,1}(1)- Y_{i,1}(0)|G_i=1,X_i\}$, as follows
\begin{equation}\label{eq: identification od did rcs}
	\operatorname{E}\{Y_{i,1}(1)- Y_{i,1}(0)|X_i,G_i=1\}=\operatorname{E}\left\{Y_i\cdot\frac{T_i-\lambda}{\lambda\,(1-\lambda)}\cdot\frac{G_i-m^G(X_i)}{m^G(X_i)[1-m^G(X_i)]}\,\Big|\, X_i\right\}\, .
\end{equation}
We then obtain the following reformulation of the null hypothesis (see the comments below):  
\begin{align}\label{eq: H0 did rcs reformulation}
	\mathcal{H}_0^{DiDrcs} \Leftrightarrow &\operatorname{E}\{[Y_{i,1}(1)-Y_{i,1}(0)]\cdot G_i|X_{c,i}\}=\operatorname{E}\{G_i|X_{c,i}\}\delta_0\nonumber \\
	\Leftrightarrow& \operatorname{E}\{\,\operatorname{E}\{[Y_{i,1}(1)-Y_{i,1}(0)]\cdot G_i|X_i\}\,|X_{c,i}\}=\operatorname{E}\{G_i|X_{c,i}\}\delta_0\nonumber\\
	\Leftrightarrow& \operatorname{E}\left\{\,\operatorname{E}\{ Y_{i,1}(1)-Y_{i,1}(0)|X_i,G_i=1\} \cdot m^G(X_i) \;| \; X_{c,i} \right\}=\operatorname{E}\{G_i|X_{c,i}\}\delta_0\nonumber\\
	\overset{(a)}{\Leftrightarrow} &\operatorname{E}\left\{ \operatorname{E}\left\{Y_i\cdot\frac{T_i-\lambda}{\lambda\,(1-\lambda)}\cdot\frac{G_i-m^G(X_i)}{m^G(X_i)[1-m^G(X_i)]}\,\Big|\, X_i\right\}\cdot m^G(X_i)\;\Big|\; X_{c,i}\right\}=\operatorname{E}\{G_i|X_{c,i}\}\delta_0\nonumber\\
	\Leftrightarrow& \operatorname{E}\left\{  Y_i\cdot\frac{T_i-\lambda}{\lambda\,(1-\lambda)}\cdot\frac{G_i-m^G(X_i)}{1-m^G(X_i)} \;\Big|\; X_{c,i} \right\}=\operatorname{E}\{G_i|X_{c,i}\}\delta_0\nonumber\\
	\Leftrightarrow& \operatorname{E}\left\{  Y_i\cdot\frac{T_i-\lambda}{\lambda\,(1-\lambda)}\cdot\frac{G_i-m^G(X_i)}{1-m^G(X_i)}- \delta_0 G_i \;\Big|\; X_{c,i}\right\}=0\, ,
\end{align}
where the equivalence $(a)$ follows from Equation \eqref{eq: identification od did rcs}. Since 
\begin{equation*}
	\operatorname{E}\{U^{DiD rcs}|X\}=\operatorname{E}\left\{  Y\cdot\frac{T-\lambda}{\lambda\,(1-\lambda)}\cdot\frac{G-m^G(X)}{1-m^G(X)}- \delta_0 G \;\Big|\; X\right\}
\end{equation*}
and $\operatorname{E}\{U^{DiD rcs}|X_c\}=0\Leftrightarrow \operatorname{E}\{U^{DiD rcs}\varphi(t^T X_c)\}=0\,\forall t\in\mathcal{T} $, Equation \eqref{eq: equivalent null hypothesis in DiD rcs} follows.   

\section{Simulations for ATE}
\label{sec: ate simulations}
We test the performance of our bootstrap test, the feasible and oracle, against the best linear predictor method (BLP) in \citet{chernozhukov2023fisherschultzlecturegenericmachine}. We propose two different DGPs, and test each method with different ML algorithms.

\subsection{Data-generating Processes}
In these simulations we test two different designs, to better illustrate the performance of different ML estimators. For each design, we generate $\mu_0(X)$ and $m^D(X)$ as particular functions of the covariates, and construct $$\mu_1(X)=(1+\gamma)\mu_0(X),$$ for $\gamma\in[0,1]$. We take $Y(1)=\mu_1(X)+\varepsilon(1)$ and $Y(0)=\mu_0(X)+\varepsilon(0)$, with $\varepsilon(d)\sim\mathcal{N}(0,1)$, independently across $d=0,1$. Hence, for $\gamma=0$, there is no treatment effect heterogeneity ($H_0$ holds), and as $\gamma$ increases we deviate from $H_0$. Moreover, we consider $D|X\sim\operatorname{Bern}\left(m^D(X)\right).$\\

\noindent\underline{Design 1:}
We first test a sparse model, so algorithms such as lasso should perform well. The model is largely based in the second model
in \citet{kennedy2023towards}. We here consider
\begin{align*}
   &X\sim N(0,I_{10});\\ 
   &\mu_0(X)=\frac{X_1+X_2}{\sqrt{2}};\\
   &m^D(X)=\left(1+\exp\left(\frac{X_1+X_2}{8\sqrt{2}}\right)\right)^{-1}
\end{align*}\\

\noindent\underline{Design 2:}
We then test a second model, based in \citet{tabord2023stratification}, that is close to sparsity, but presents non-linearities. Other methods, such as random forest, should perform better under this DGP. Here we take, 
\begin{align*}
    &X_j\sim\operatorname{Beta}(2,5) \text{ independently across } j=1,\ldots,10;\\
    &\mu_0(X)=\sum_{j=1}^{10} \mathbbm{1}\{X_j>0.4\}X_j^2\times(-1)^{j-1}\times10^{2-j};\\
    &m^D(X)=0.4+\sum_{j=1}^{2}\times\mathbbm{1}\{X_j>0.1\}\frac{X_j}{6}.
\end{align*}

\subsection{Performance}
We evaluate the rejection rate under the null and alternative hypotheses over 1,000 simulations. We consider sample sizes $n\in\{500,1000\}$, with $B=499$ bootstrap iterations. We test for different subsets $X_c$ containing the first $d\in\{2,5\}$ components of $X$, i.e., $X_c=(X_1,X_2)$ or $X_c=(X_1,\ldots,X_5)$. As ML algorithms for the nuisance functions, we use lasso and random forests. Cross-fitting is performed with five equally sized random folds. Moreover, we set $\phi_t(\cdot)=\exp(\bm{i}t^\top\cdot)$ and let $\rho(\cdot)$ be the Gaussian density, so that $\mathcal{F}_\rho$ corresponds to a Gaussian kernel function with bandwidth $h=1$.

Table \ref{tb:null_hyp_ATE} shows the rejection probabilities under the null ($\gamma=0$) for $\alpha=5\%$ and $\alpha=10\%$. Notice, that the BLP performs highly conservative tests, whereas our test is almost exactly valid across different sample sizes, ML estimators and number of covariates $d$.\footnote{\cite{Imai2025b} find in simulations also conservative confidence intervals of \cite{chernozhukov2023fisherschultzlecturegenericmachine} approach.} However, increasing the number of covariates makes the test slightly more conservative.
\begin{table}[htb]
	\caption{Rejection Rate under the Null Hypothesis}
	\label{tb:null_hyp_ATE}
    \centering
    \begin{adjustbox}{max width=\textwidth, max totalheight=\textheight}
	\begin{threeparttable}
\begin{tabular}{rlSSSSSSSSSSS}   
\toprule\toprule
     &            &\multicolumn{5}{c}{$\alpha=0.05$}&  &\multicolumn{5}{c}{$\alpha=0.1$} \\
      \cmidrule(lr){3-7} \cmidrule(lr){9-13}
      &            &\multicolumn{2}{c}{Design 1}&  &\multicolumn{2}{c}{Design 2}& &\multicolumn{2}{c}{Design 1}&  &\multicolumn{2}{c}{Design 2}                      \\
      \cmidrule(lr){3-4} \cmidrule(lr){6-7}\cmidrule(lr){9-10} \cmidrule(lr){12-13}
      
 $n$  & Method     &\mc{$d=2$} &\mc{$d=5$}   &  &\mc{$d=2$} &\mc{$d=5$}&&\mc{$d=2$} &\mc{$d=5$}   &  &\mc{$d=2$} &\mc{$d=5$}   \\\midrule
      & FICM Lasso & 0.050 & 0.045 & 0.038 & 0.028 && 0.105 & 0.088 & 0.076 & 0.067\\ 
      & FICM RF    & 0.037 & 0.034 & 0.046 & 0.033 && 0.096 & 0.087 & 0.095 & 0.069\\
500   & OICM       & 0.046 & 0.041 & 0.040 & 0.036 && 0.097 & 0.088 & 0.091 & 0.082\\ 
      & BLP Lasso  & 0.005 & \text{--} & 0.009 & \text{--} && 0.013 & \text{--} & 0.021 & \text{--}\\   
      & BLP RF     & 0.001 & \text{--} & 0.001 & \text{--} && 0.002 & \text{--} & 0.009 & \text{--}\\\midrule
      & FICM Lasso & 0.050 & 0.048 & 0.041 & 0.030 && 0.090 & 0.084 & 0.081 & 0.078\\
      & FICM RF    & 0.042 & 0.042 & 0.049 & 0.034 && 0.081 & 0.095 & 0.108 & 0.081\\
1000  & OICM       & 0.050 & 0.046 & 0.047 & 0.037 && 0.094 & 0.090 & 0.100 & 0.080\\
      & BLP Lasso  & 0.006 & \text{--} & 0.006 & \text{--} && 0.015 & \text{--} & 0.017 & \text{--}\\
      & BLP RF     & 0.003 & \text{--} & 0.003 & \text{--} && 0.011 & \text{--} & 0.009 & \text{--}\\ 
   \bottomrule\bottomrule
\end{tabular}
\begin{tablenotes}[flushleft,para]
\footnotesize
This table reports rejection rates across 1,000 Monte Carlo simulations under the null hypothesis for the bootstrapped test for ATE under unconfoundedness. The sample size is $n\in\{500, 1000\}$. Nuisance functions are estimated using either lasso or random forests. Cross-fitting is performed with five folds. For each Monte Carlo iteration, we perform 499 bootstrap replications. We report rejection rates for significance levels $\alpha\in\{5\%,10\%\}$ and for different dimensions of the vector $X_c$ ($d=2,5$). The Fourier matrix is computed using a Gaussian kernel with bandwidth $1$. We compare the results of our test to the case where we use the Oracle statistic and to the BLP test in \citet{chernozhukov2023fisherschultzlecturegenericmachine}, with lasso and random forests.
\end{tablenotes}
\end{threeparttable}
\end{adjustbox}
\end{table}

\begin{figure}[htb]
    \centering
    \includegraphics[width=\linewidth]{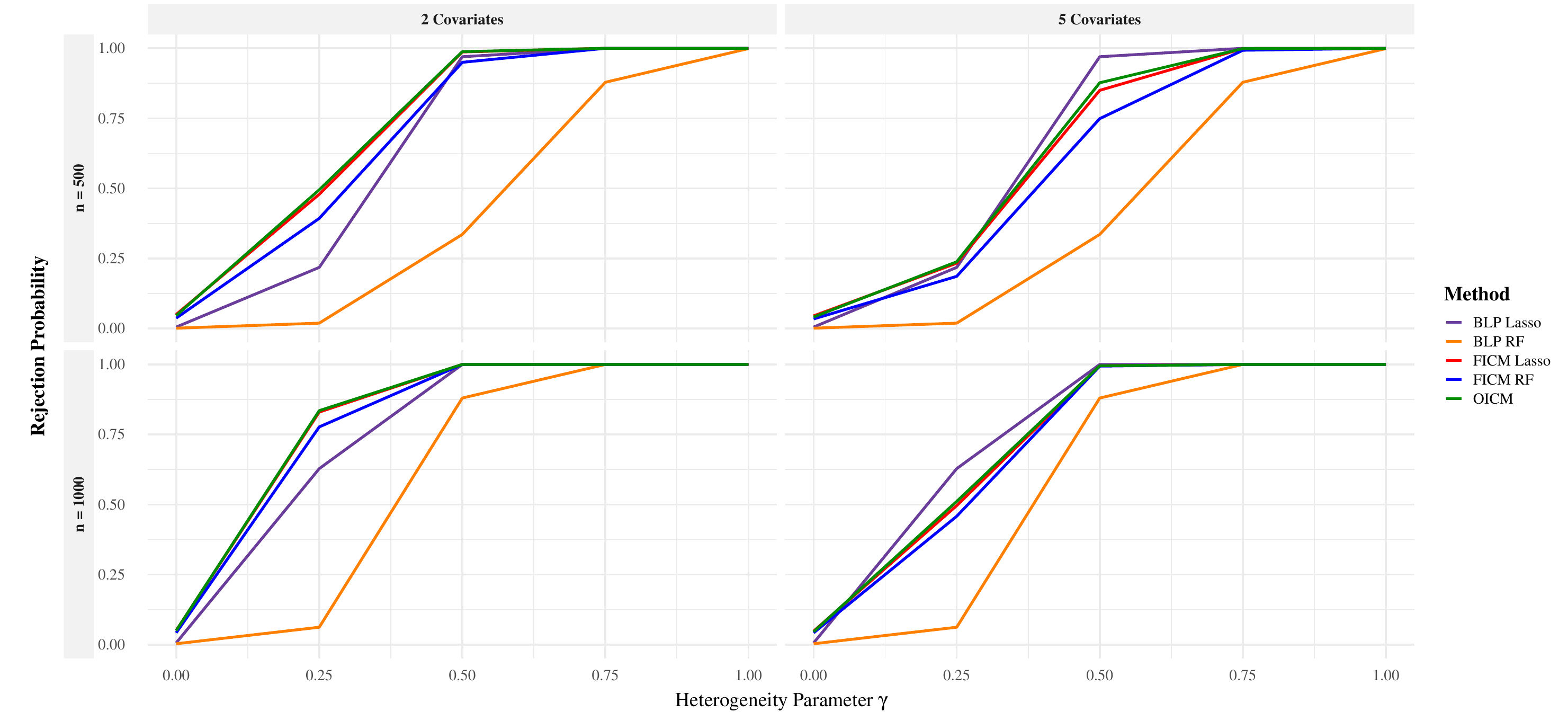}
    \caption{Rejection Rates under Design 1}
    \label{fig:Kennedy}
     \floatfoot{\justifying This figure shows rejection in Design 1 rates of 95\% confidence intervals for different values of $\gamma$, which control the heterogeneity of the ATE, for the bootstrapped test in the ATE under unconfoundeness. Rejection rates are computed from 1,000 Monte Carlo simulations. The sample size is $n\in\{500,1000\}$. Nuisance functions are estimated using either lasso or random forests. Cross-fitting is performed with five folds, and 499 bootstrap replications are used in each iteration. We consider different dimensions of the vector $X_c$ ($d=2,5$). The Fourier matrix is computed using a Gaussian kernel with bandwidth $1$. We compare the results of our test to the case where we use the Oracle statistic and to the BLP test in \citet{chernozhukov2023fisherschultzlecturegenericmachine}, with lasso and random forests.}
\end{figure}

\begin{figure}[htb]
    \centering
    \includegraphics[width=\linewidth]{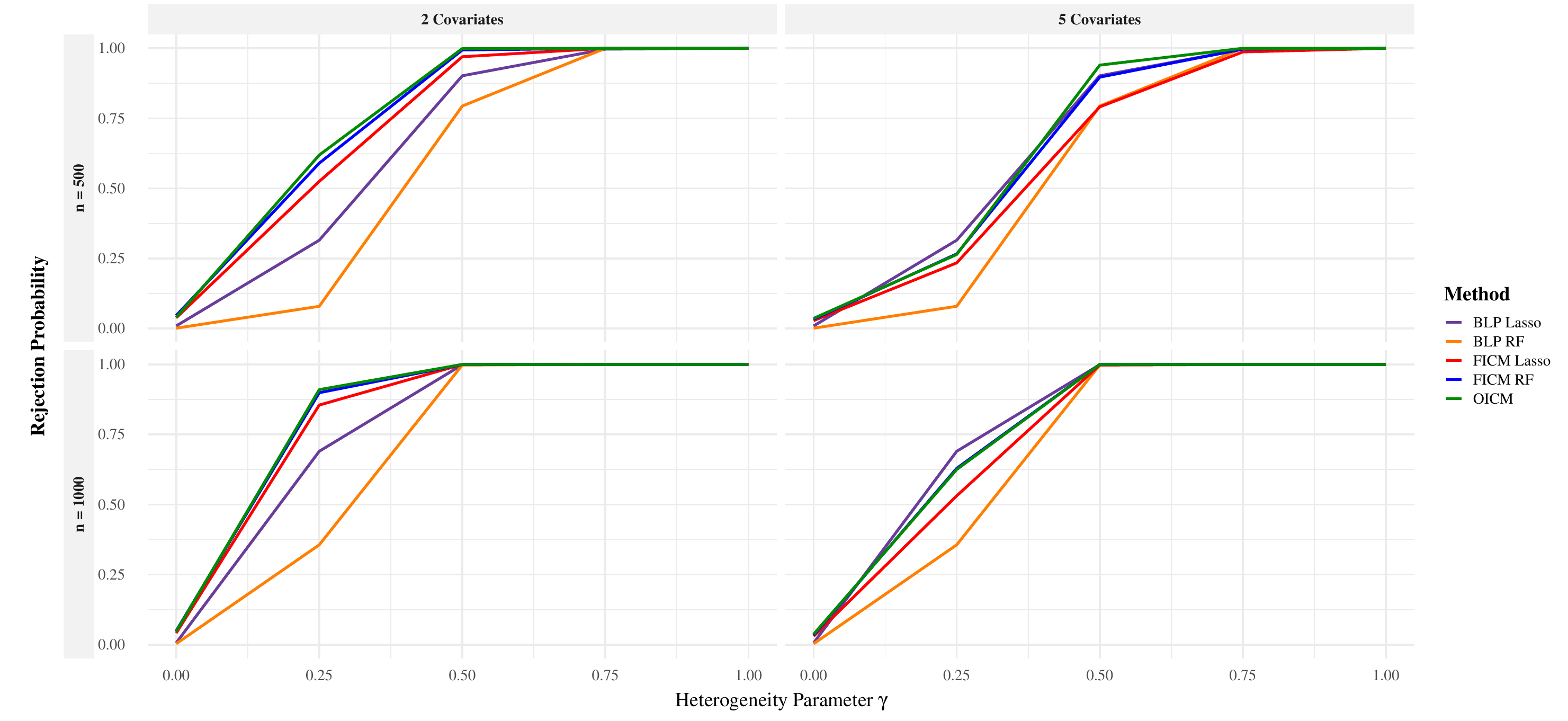}
    \caption{Rejection Rates under Design 2}
    \label{fig:TM}
    \floatfoot{\justifying This figure shows rejection in Design 2 rates of 95\% confidence intervals for different values of $\gamma$, which control the heterogeneity of the ATE, for the bootstrapped test in the ATE under unconfoundeness. Rejection rates are computed from 1,000 Monte Carlo simulations. The sample size is $n\in\{500,1000\}$. Nuisance functions are estimated using either lasso or random forests. Cross-fitting is performed with five folds, and 499 bootstrap replications are used in each iteration. We consider different dimensions of the vector $X_c$ ($d=2,5$). The Fourier matrix is computed using a Gaussian kernel with bandwidth $1$. We compare the results of our test to the case where we use the Oracle statistic and to the BLP test in \citet{chernozhukov2023fisherschultzlecturegenericmachine}, with lasso and random forests.}
\end{figure}

Figures \ref{fig:Kennedy} and \ref{fig:TM} show the power curves for different values of $\gamma$. We see that, as predicted lasso performs  better under Design 1 and random forests under Design 2. Moreover, we see that in general, the rejection of our tests is greater than when using the BLP method in \citet{chernozhukov2023fisherschultzlecturegenericmachine}, especially when there is a smaller level of heterogeneity. Notably, the bootstrap test is less sensible to the ML method used to estimate the nuisance functions.

\end{document}